\documentclass[12pt]{article}
\usepackage[body={17.5cm, 22cm},right=2cm]{geometry}
\usepackage{amssymb,graphicx}
\usepackage{amsmath}
\usepackage{epsfig}
\usepackage{hyperref}

\def\d{\partial}
\def\l{\left(}
\def\r{\right)}

\newenvironment{changemargin}[2]{%
  \begin{list}{}{%
    \setlength{\topsep}{0pt}%
    \setlength{\leftmargin}{#1}%
    \setlength{\rightmargin}{#2}%
    \setlength{\listparindent}{\parindent}%
    \setlength{\itemindent}{\parindent}%
    \setlength{\parsep}{\parskip}%
  }%
  \item[]}{\end{list}}

\newcommand{\be}{\begin{equation}}
\newcommand{\ee}{\end{equation}}
\newcommand{\bea}{\begin{eqnarray}}
\newcommand{\eea}{\end{eqnarray}}
\newcommand{\bg}{\begin{gather}}
\newcommand{\eg}{\end{gather}}
\newcommand{\bseq}{\begin{subequations}}
\newcommand{\eseq}{\end{subequations}}

\newcommand{\Tr}{{\rm Tr}}

\begin{document}
\begin{center}
\begin{changemargin}{-0.5cm}{-0.5cm}
{\Large\bf Exploring the String Axiverse with  Precision Black Hole Physics}\\
\end{changemargin}
\vspace{0.5cm}
{ \large
Asimina Arvanitaki$^{a,b}$ and
 Sergei~Dubovsky$^{c,d}$\\ 
\vspace{.45cm}
{\small  \textit{  $^{\rm a}$ Berkeley Center for Theoretical Physics, University of California, Berkeley, CA, 94720}}\\ 
\vspace{.1cm}
{\small  \textit{  $^{\rm b}$ Theoretical Physics Group, Lawrence Berkeley National Laboratory, Berkeley, CA, 94720 }}\\ 
\vspace{.1cm}
{\small  \textit{  $^{\rm c}$ Department of Physics, Stanford University, Stanford, CA 94305, USA }}\\ 
\vspace{.1cm}
{\small  \textit{  $^{\rm d}$
Institute for Nuclear Research of the Russian Academy of Sciences, 
        60th October Anniversary Prospect, 7a, 117312 Moscow, Russia}}\\}
\end{center}
\begin{center}
\begin{abstract}
It has recently been suggested that the presence of a plenitude of light axions, an Axiverse, is evidence for the extra dimensions of string theory. We discuss the observational consequences of these axions on astrophysical black holes through the Penrose superradiance process. When an axion Compton wavelength is comparable to the size of a black hole, the axion binds to the black hole ``nucleus" forming a gravitational atom in the sky. The occupation number of superradiant atomic levels, fed by the energy and angular momentum of the black hole, grows exponentially. The black hole spins down and an axion Bose-Einstein condensate cloud forms around it. When the attractive axion self-interactions become stronger than the gravitational binding energy, the axion cloud collapses, a phenomenon known in condensed matter physics as ``Bosenova". The existence of axions is first diagnosed by gaps in the mass vs spin plot of astrophysical black holes.
For young black holes the allowed values of spin are quantized, giving rise to "Regge trajectories" inside the gap region. The axion cloud can also be observed directly either through precision mapping of the near horizon geometry or through gravitational waves coming from the Bosenova explosion, as well as axion transitions and annihilations in the gravitational atom. Our estimates suggest that these signals are detectable in upcoming experiments, such as Advanced LIGO, AGIS, and LISA. Current black hole spin measurements imply an upper bound on the QCD axion decay constant of $2\cdot 10^{17}$ GeV, while Advanced LIGO can detect signals from a QCD axion cloud with a decay constant as low as the GUT scale. We finally discuss the possibility of observing the $\gamma$-rays associated with the Bosenova explosion and, perhaps, the radio waves from axion-to-photon conversion for the QCD axion.\end{abstract}
\end{center}
\newpage
\tableofcontents
\section{Introduction and summary}
Black holes are among the most fascinating systems in astrophysics, and the most mysterious objects in quantum gravity and string theory, for a long time serving as  principal ``theoretical laboratories"  for exploring non-perturbative gravitational dynamics. The purpose of  this paper is to initiate a detailed study of the exciting possibility~\cite{Arvanitaki:2009fg} that astrophysical black holes may serve as  {\it actual}   laboratories for the discovery of new elementary particles.

There are several reasons why we believe this possibility is realistic.
On a purely phenomenological side, black hole observations are routine practice in nowaday astronomy (see, e.g.,  \cite{Narayan:2005ie} for a review). About 40 stellar mass black holes in X-ray binaries in the Milky Way and neighboring galaxies have been identified with masses in the range $\sim 5\div 20 M_\odot$. Supermassive black holes with masses $\sim 10^5\div 10^{10}~M_{\odot}$ have been found in centers of many galaxies including the Milky Way   and believed to be hosted by nearly all of the galaxies.
Also, the first intermediate mass ($\sim 100\div 10^{5}~M_{\odot}$) candidates have been identified.

Following the evolution of binary systems or measuring the velocity dispersion of stars rotating around  galactic centers allows to determine black hole masses. Most crucially for what follows, recent advances in X-ray astronomy and in numerical magnetohydrodynamical simulations of the accreting gas in the Kerr metric open the possibility for  a detailed exploration of  the near-horizon region and, as a consequence, for high precision neasurements of black 
hole spins~\cite{McClintock:2009as,Brenneman:2009hs}. First estimates for the angular momentum of several black holes have already been delivered~\cite{McClintock:2009dn}, often suggesting high values for the spin, although at the moment different techniques sometimes give rise to conflicting results \cite{Blum:2009ez}.

In the future, apart from improvements of traditional astronomical techniques for observing the near horizon environment and its better theoretical modeling, a unique probe of the black hole geometry will be provided by low frequency gravitational waves observatories, such as LISA \cite{LISA1} or  AGIS, a gravitational wave detector based on atom interferometric techniques \cite{Dimopoulos:2008sv,Dimopoulos:2007cj}. For the purpose of testing the near horizon geometry the most promising candidates are the so-called extreme mass ratio inspirals---stellar mass compact objects captured by supermassive black hole in the galactic center (see, e.g., \cite{Hughes:2006pm}). LISA and AGIS are expected to detect about a hundred of such events per year.  Each such measurement allows not only to determine the black hole spin and mass with an exquisite accuracy, $10^{-3}\div 10^{-5}$, depending on the details of a particular event, but also to check whether higher order  metric moments, up to $6\div 7$, agree with their values for the Kerr geometry.

This ongoing observational progress indicates that we are witnessing the dawn of  precision black hole physics. Undoubtedly, black hole observations will be of great value for astrophysics, however it is natural to inquire whether these data may be useful for beyond the Standard Model physics as well, given that  it will provide a rare test of non-linear gravity. However, possibly contrary to naive expectation, it turns out quite challenging to find modified gravity theories which would predict deviations from general relativity near astrophysical black holes and would not contradict current gravity tests. One candidate class of modified theories of gravity affecting black hole dynamics are models of Higgs phases of gravity, where black hole no-hair theorems can be violated~\cite{Dubovsky:2007zi}.

In this paper we explore a less exotic possibility  to test fundamental physics with precision black hole observations.
It is related to the famous Penrose process, a mechanism to extract energy and angular momentum from rotating black holes \cite{Penrose:1969pc,Christodoulou:1970wf}. As reviewed in detail below, this process, known as superradiance, when applied to waves rather than particles  \cite{Zeld,Misner:1972kx,Starobinskii}, gives rise to a spin-down instability of a rotating black hole \cite{Damour:1976kh,Ternov:1978gq,Zouros:1979iw,Detweiler:1980uk}, if a massive boson with a Compton wavelength of order the black hole gravitational radius is present in nature. As we will see, this instability turns rotating astrophysical black holes into sensitive detectors of bosons with masses in the range $\mu\sim 10^{-9}\div 10^{-21}$~eV. Before focusing on the observational consequences of the superradiant instability, let us review why it is natural to expect ultra-light bosons in the theory that transform astrophysical black holes in probes of fundamental physics.

A natural situation giving rise to  a particle of a small, but non-vanishing mass is when this particle is a (pseudo)Goldstone boson of a spontaneously broken global symmetry, which is also explicitly broken by non-perturbative effects. Probably the best motivated candidate for such a particle is the QCD axion $\phi_a$---a pseudoscalar particle coupled to the QCD instanton number density via
\be
\label{axionaction}
S_\theta={1\over 32\pi^2f_a}\int d^4x\;\phi_a\epsilon^{\mu\nu\lambda\rho}\Tr \,G_{\mu\nu}G_{\lambda\rho} \, .
\ee
Note that at the classical level $S_\theta$ is invariant under the
 Peccei--Quinn (PQ) symmetry $\phi_a\to \phi_a+const$, so that the QCD axion is indeed a (pseudo)Goldstone boson with $f_a$ being the scale of spontaneous symmetry breaking. This symmetry is explicitly broken by the QCD instanton effects that generate the axion potential giving rise to a solution for the strong CP problem---the primary motivation for the QCD axion.
 As a result the QCD axion acquires a mass equal to 
\be
\label{QCDmass}
\mu_a\approx 6\cdot 10^{-10} \mbox{eV} \l{10^{16}\mbox{GeV}\over f_a}\r  \;.
\ee
The Compton wavelength of the QCD axion with a high symmetry breaking scale $f_a\gtrsim 10^{16}$~GeV matches the size of stellar mass black holes and, consequently, can affect their dynamics, suggesting that this part of the parameter space for the QCD axion can be explored through black hole observations. 

There are several reasons why this conclusion is very important. First, non-gravitational interactions of the QCD axion with the rest of the Standard Model particles are very suppressed at these high values of $f_a$. As a result this part of the parameter space can not be easily probed by any other means, either laboratory or astrophysical. Second, in many ``generic" string constructions, i.e., in compactifications where the extra-dimensional manifold is neither highly anisotropic, nor strongly warped, the values of $f_a$ are naturally around the grand unification scale $M_{GUT}\simeq2\times 10^{16}$~GeV
\cite{Svrcek:2006yi}. Finally, as elaborated in more detail in section~\ref{anthropic}, finding the QCD axion with $f_a\sim M_{GUT}$ would indicate that the baryon-to-dark matter ratio varies on length scales longer than the observed part of the Universe and its local value is determined by anthropic reasoning. Discovery of the QCD axion in this regime would be further evidence for enviromental selection already suggested by the cosmological constant problem, and  by the string landscape.


There is an even stronger and more direct link between the QCD axion and the landscape of string vacua, a link that gives rise to the expectation of a {\it plenitude} of light axion-like particles, an axiverse~\cite{Arvanitaki:2009fg} -- this same link also suggests the existence of many massless vectors, whose massive superpartners may be discovered at the LHC~\cite{Arvanitaki:2009hb}.
In string constructions, an axion usually arises as a Kaluza--Klein (KK) zero mode of a higher-dimensional  antisymmetric form field.
Such zero modes have a purely topological origin: they are labeled by non-contractable cycles in the extra-dimensional manifold. Non-contractable cycles allow for non-trivial gauge field configurations with a vanishing field strength, the so called Wilson lines. These configurations do not carry energy and correspond to zero KK modes at the perturbative level. They only acquire a mass due to non-perturbative effects.

Interestingly, the very same ingredients that give rise to the string axiverse, higher-dimensional form fields and non-trivial cycles in the compactification manifold allowing also to turn on gauge fluxes, also give rise to the string landscape of $10^{500}$ or so vacua. 
In order to allow for the tuning of the cosmological constant at the $\sim 10^{-120}$ level, as required by observations, the compactification manifold should contain of order few hundred cycles, given that the total amount of flux quanta for a cycle is typically limited by a number around ten in order to stay in a  perturbative regime. Consequently, one may expect hundreds of axion-like particles in a given string compactification. However, a plenitude of cycles does not yet guarantee the presence of a plenitude of axions.  There is a number of effects in string theory that could produce a large axion mass, such as branes wrapping the cycles, and fluxes. One can roughly estimate the number of light axions as being determined by the number of cycles without fluxes---presumably, around one tenth of the total number of cycles. Still this leaves us with the expectation of several tens of axion-like particles. 

The discovery of a plenitude of particles in {\it our vacuum} with similar properties but different masses supports the idea of a plenitude of {\it vacua}, as both the axiverse and the multiverse are dynamical consequences of the same fundamental ingredients.

The masses of string axions are exponentially sensitive to the sizes of the corresponding cycles, so one expects them to be homogeneously distributed on the logarithmic scale. However, given that the QCD $\theta$-parameter is constrained to be less than $10^{-10}$, non-perturbative string corrections to the QCD axion potential should be at least ten orders of magnitude suppressed as compared to the QCD generated potential. It is then natural to expect many of the axions to be much lighter than the QCD axion; these are the axions whose mass is dominated only by these small non-perturbative string effects.

The implicit, and very plausible assumption behind this line of reasoning is that there is no anthropic reason for the existence and properties of the QCD axion. Consequently, these properties should follow from the dynamics of the compactification manifold, rather than being a result of fine-tuning, and the QCD axion should be a typical representative among other axion-like fields. A priori we expect tens (or even hundreds) of light axions, it would be really surprising if the QCD axion turned out to be the single one.
\begin{figure}[t!] 
 \begin{center}
 \includegraphics[width=5in]{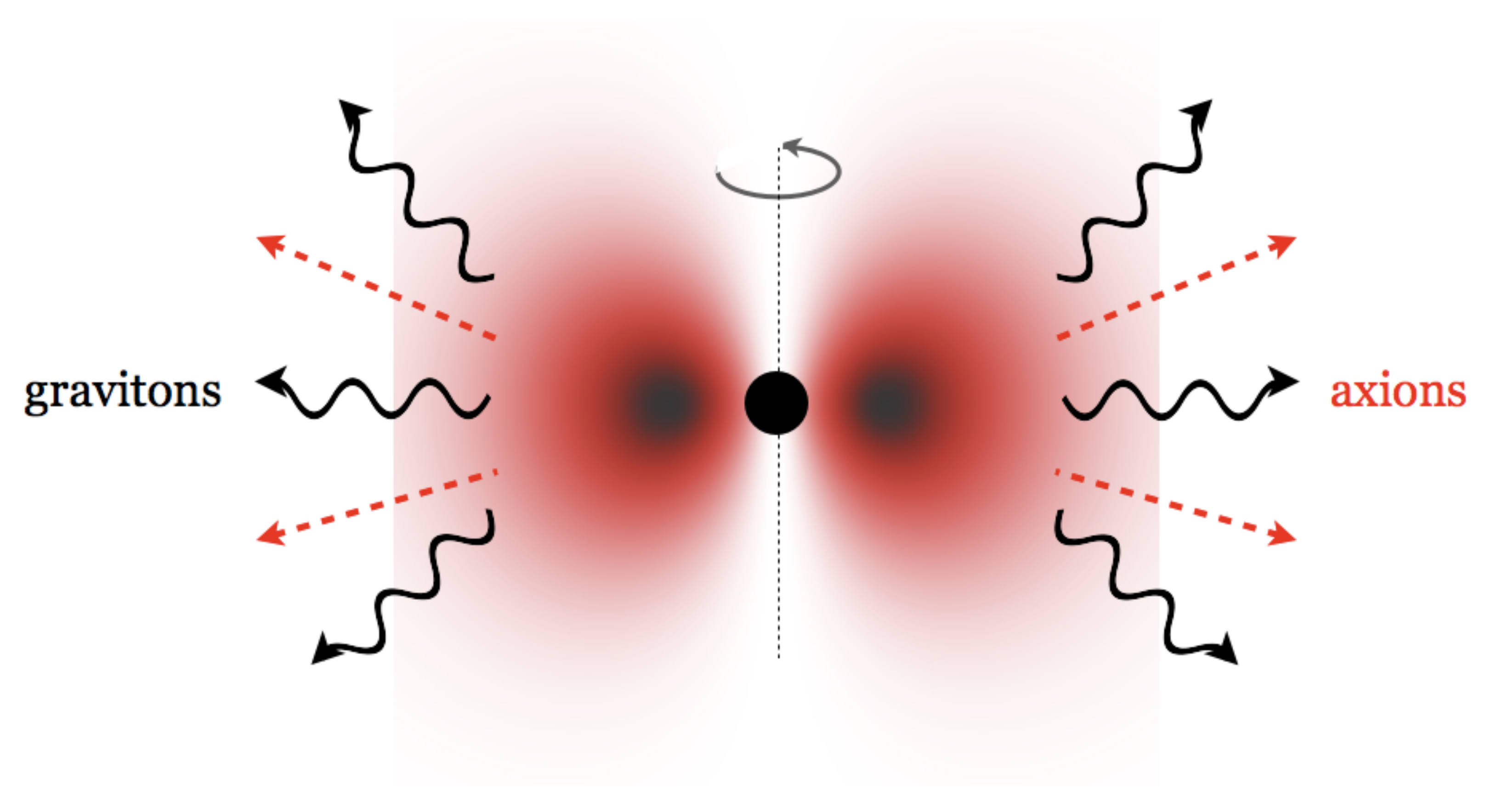}
 \caption{{\bf Axionic Black Hole Atom:} The spinning black hole ``feeds" superradiant states forming an axion Bose-Einstein condensate. 
 The resulting bosonic atom will emit gravitons through axion transitions between levels and annihilations and will emit axions  as a consequence of self-interactions in the axion field.}
 \label{carnotcycle}
 \end{center}
\end{figure}

These arguments motivate us to look not only for the QCD axion, but for axions in the entire mass range  $\mu\sim 10^{-9}\div 10^{-21}$~eV, where they can affect stellar or galactic astrophysical black holes through superradiance. Let us summarize now the major features of superradiance and its principal observational consequences. 

Superradiance \cite{Zeld,Misner:1972kx,Starobinskii} is the phenomenon of wave amplification during scattering off a rotating black hole which takes place whenever the wave frequency $\omega$ and the magnetic quantum number $m$ satisfy the superradiance condition
\be
\label{Omegacond}
0<\omega<mw_+\;,
\ee 
where $w_+$ is the angular velocity of the black hole horizon defined as
\be
w_+ = \frac{1}{2r_g} \frac{{a}/{r_g}}{1+\sqrt{1 - (a/r_g)^2}}
\label{spinvelo}
\ee
with $r_g$ being the gravitational radius of the black hole and $0<a<r_g$ is the spin-to-mass ratio. Superradiant amplification may lead to an instability under certain circumstances. 
One example, admittedly not a very practical one, is the ``black hole bomb"  by Press and Teukolsky \cite{Press:1973zz,Press:Nature}: a rotating black hole surrounded by a spherical mirror.
A single photon introduced in the system, or created by quantum fluctuations, with quantum numbers satisfying the superradiance condition (\ref{Omegacond}) gives rise to an exponentially growing number of photons inside the mirror through a chain of consequent amplifications at the black hole horizon and reflections from the mirror. 

Remarkably, nature provides such a mirror in the presence of a massive boson  \cite{Damour:1976kh,Ternov:1978gq,Zouros:1979iw,Detweiler:1980uk}. 
Massive bosons, in our case axions, have bound Keplerian levels in the gravitational field of a black hole.
This allows for a black hole to release its spin by populating levels satisfying the condition (\ref{Omegacond}), see Fig.~\ref{carnotcycle}.
This creates an axionic Bose-Einstein condensate (BEC) cloud rotating around the black hole. The process is only efficient if the Compton wavelength of an axion is comparable to the black hole size. As a result, the energy spectrum of superradiant levels is quantized and very close to the 
spectrum of a hydrogen atom. The superradiant axion cloud loses its energy and momentum by gravitational wave emission associated with axion transitions between different  ``atomic" levels, and with axion annihilations to gravitons. Another important loss mechanism is related to non-linearities in the axion potential and results in the emission of axions. Finally, the whole system may be fueled by energy and spin inflow from the matter accreting onto the black hole. 

\begin{figure}[t] 
 \begin{center}
 \includegraphics[width=6in,trim=70 50 50 100]{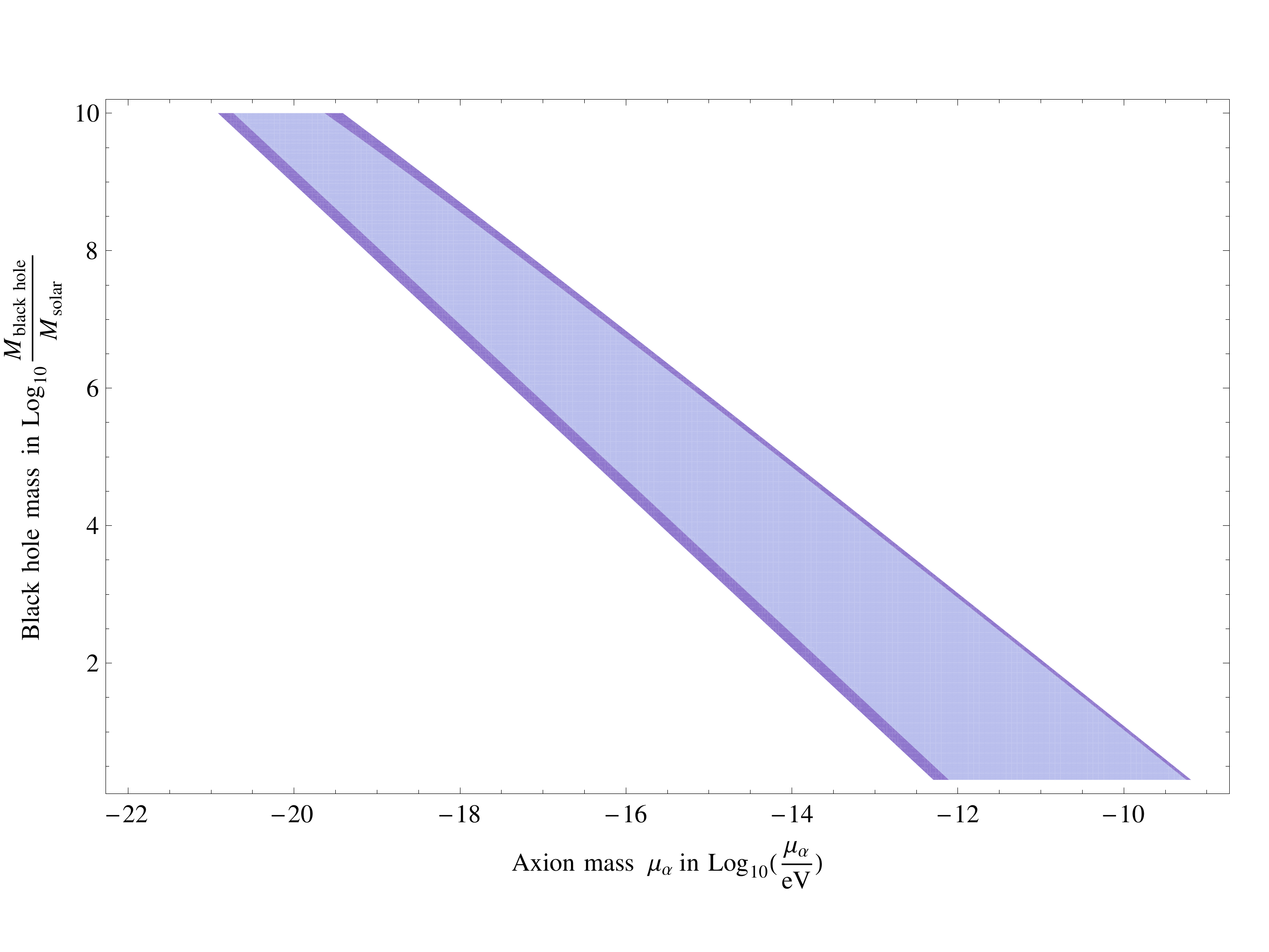}
 \caption{The part of the black hole and axion parameter space potentially affected by superradiance. For axion and black hole masses in the colored region the time required  to create a substantial axion cloud is shorter than the age of the Universe. For masses in a light colored region the superradiance rate is faster than the Eddington accretion rate.
 }
 \label{aeq1plot}
 \end{center}
\end{figure}
In Fig. \ref{aeq1plot} we show the region of black hole mass and axion mass parameter space that is affected by superradiance.
The dark area outlines the region where a superradiant cloud has enough time to be built up during the lifetime of the Universe for a maximally spinning black hole. The lighter region shows the part of the parameter space where the superradiant spindown rate is faster than the spinup rate due to Eddington accretion, so that Eddington accreting black holes in this mass range lose their spin as the cloud develops.

There are three major classes of observational signatures associated with the axion cloud that we are going to discuss in the current paper:
\begin{itemize}
\item Gaps in the black hole ``Regge plot" (spin vs mass plane)---the absence of rapidly rotating black holes when their gravitation radius matches the Compton wavelength of an axion.
\item Direct gravitational wave signals from the rotating axion cloud. For the QCD axion this signal falls into the sensitivity band of the Advanced LIGO interferometer.
\item Modification of the near-horizon metric due to the presence of the axion cloud.
\end{itemize}
The main goal of this paper is to evaluate the observability of these classes of signatures in near future experiments. Detailed quantitative predictions for gravitational wave emission rates and waveforms, as well as the change of the templates for the near horizon metric is likely to require numerical work. The reason is that the axion BEC cloud is a rich and complicated dynamical system with many processes that need to be taken into account in order to have an accurate description of its behavior at cosmological time scales. 
One important process that poses a challenge for an analytical treatment is the effect of axion self-interactions on  superradiance. Here we limit ourself to a qualitative, semi-analytical discussion of the major physical processes involved. Our analysis indicates that depending on the relation between the axion mass and the black hole size all three classes of signatures can be observable for different systems. This  provides strong motivation for further numerical analysis of axionic superradiance. 

\begin{figure}[t!] 
 \begin{center}
 \includegraphics[width=5in,trim= 0 0 0 40]{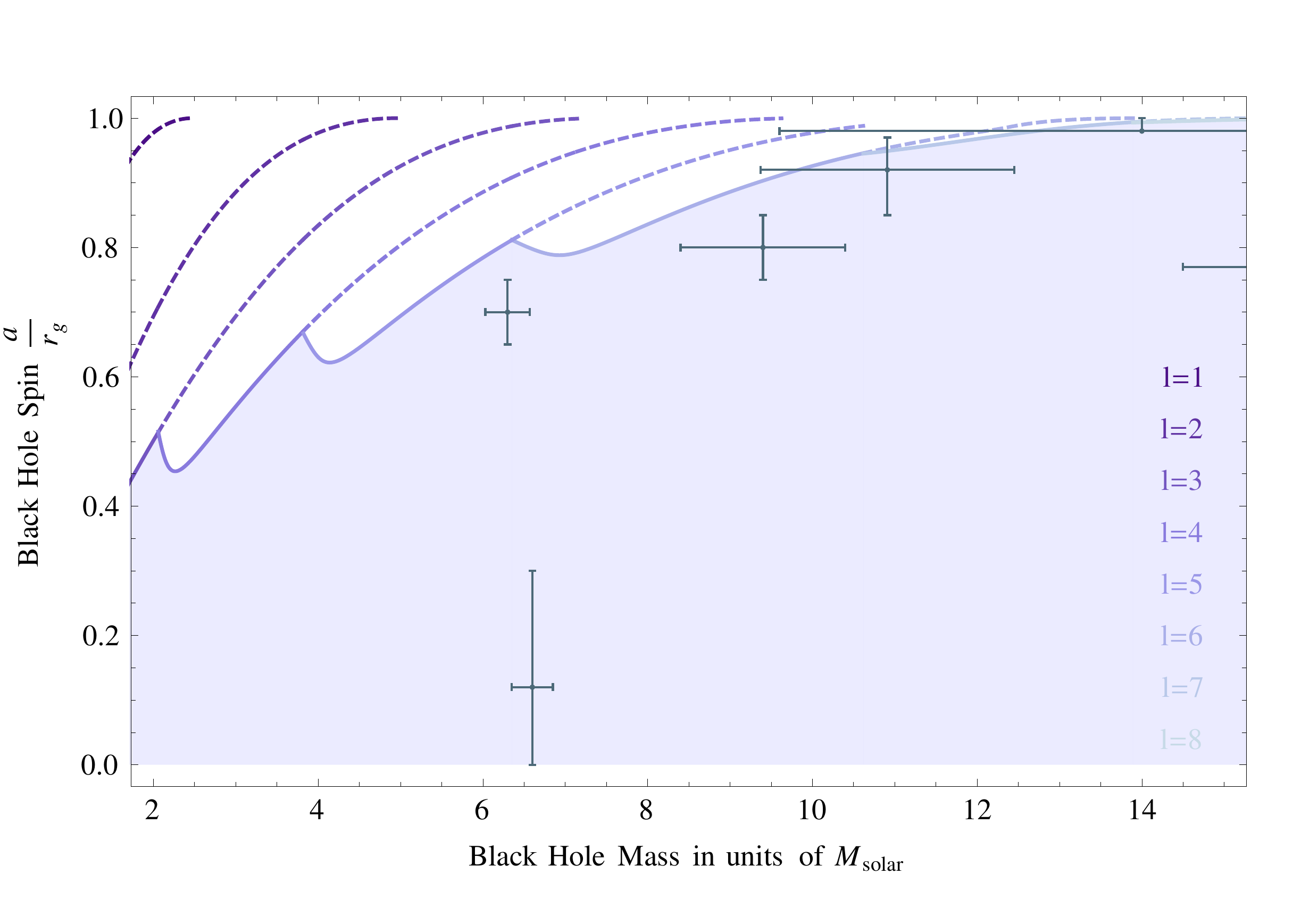}
 \end{center}
 \begin{center}
 \includegraphics[width=5in,trim= 0 0 0 70]{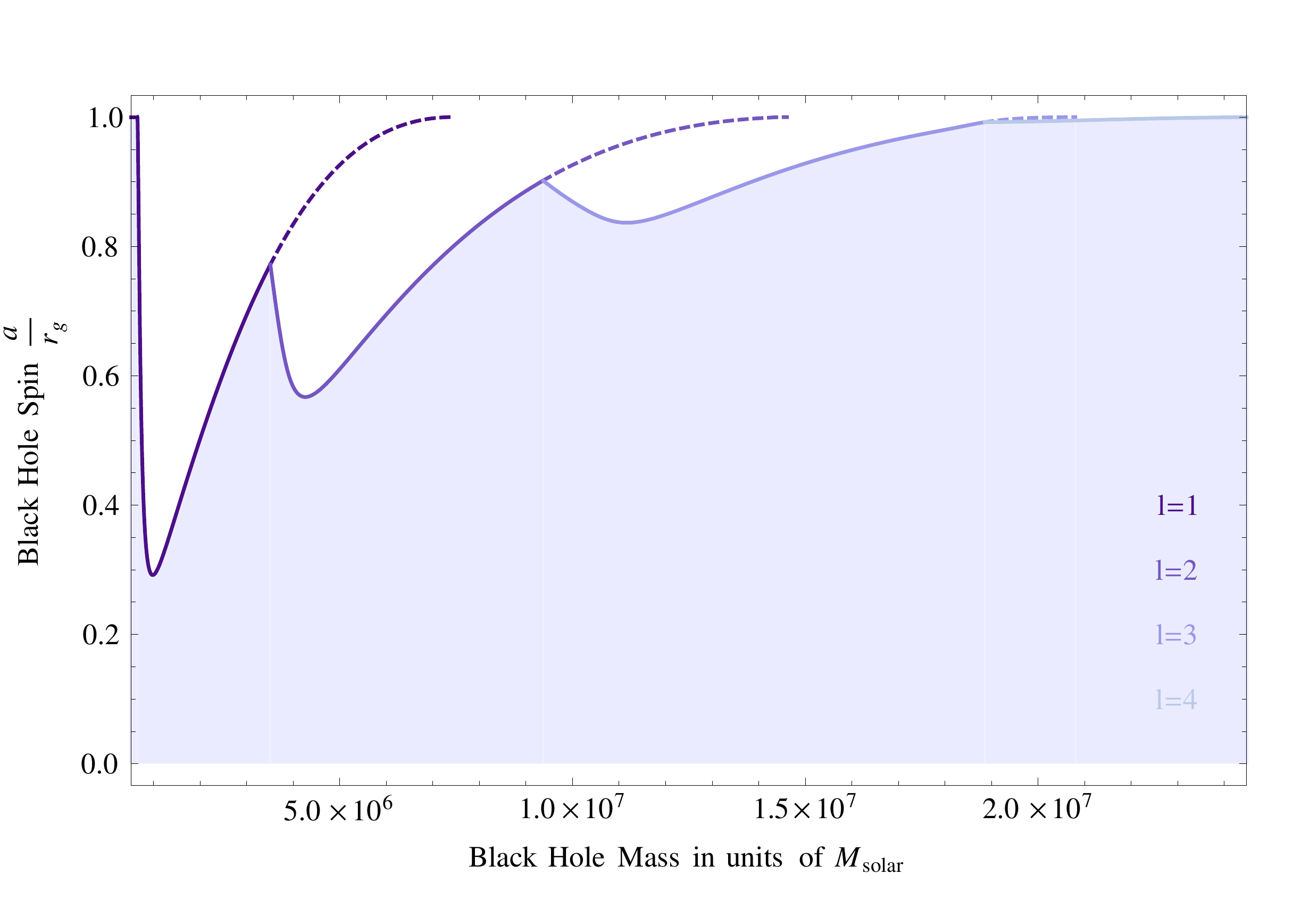}
 \caption{The regions in the black hole Regge plane affected by  superradiance for the QCD axion with $m=3\cdot 10^{-11}$~eV (the upper panel) and for a lighter axion with $m=10^{-17}$~eV. The data points correspond to spin measurements obtained by fitting the thermal continuum X-ray spectra \cite{McClintock:2009dn}. Old black holes are expected to be found in the shaded region, where they are not affected by superradiance. Young black holes may be found also on the dashed colored lines inside the gap. Different colors correspond to superradiant levels with different orbital quantum numbers $l$.
 }
 \label{fig_summary}
 \end{center}
\end{figure}

Let us briefly summarize our main results. In Fig. \ref{fig_summary} we show the black hole Regge plot for two different axion masses. The upper plot corresponds to an axion mass of $\mu_a=3\cdot 10^{-11}$ eV, which is the mass of the QCD axion for $f_a=2\cdot 10^{17}$ GeV, and the lower illustrates the effect of a lighter axion with $\mu_a=10^{-17}$ eV. Black holes in the shadowed region are not affected by superradiance during the age of the Universe. Sufficiently old black holes are expected to be found in this region. In the plot corresponding to the QCD axion we also present existing spin measurements; this data suggest an upper bound on the QCD axion decay constant of $2 \cdot 10^{17}$ GeV.

Note that the uncolored gap regions exhibit internal structure (dashed lines), reflecting the quantized behavior of the gravitational black hole atom. Black holes may stay on these lines, the ``Regge trajectories", for cosmological time scales. 
In order to understand this behavior, we should note that there is a number of different superradiant bound states labeled by different angular quantum number $l$. The instability time-scale rapidly drops down as the orbital quantum number $l$ increases. Therefore, the spin-down process of the black hole is initially driven by the level with the minimum value of $l$ for which the superradiance condition (\ref{Omegacond}) is satisfied, and stops when enough spin has been extracted, so that the superradiance instability rate goes to zero --- the superradiance condition is now saturated. At this point the black hole finds itself on one of the dashed lines of the Regge plot. 

Next, we should expect that the black hole spin-down proceeds with the growth of the level with the next-to-largest superradiance rate, i.e., the $l+1$ level. However, further spin-down is damped when the black hole reaches one of the Regge trajectories as a consequence of axion self-interactions. The axions bound to the level that is no longer superradiating serve as an axially asymmetric perturbation to the system that mixes superradiant with non-superradiant levels and shuts off the black hole spin-down process. This is similar to introducing a non-spherical defect on the mirror in the Press-Teukolsky black hole bomb: photons reflected from the mirror exit the superradiant region (\ref{Omegacond}) and are now absorbed by the black hole, turning the instability off.
As a result of these axion non-linearities, black holes stay on the Regge trajectories until various axion loss processes, or a violent accretion event, dissipate the cloud down to a small enough size, such that non-linearities no longer inhibit the instability. Then the black hole rapidly jumps to the next Regge trajectory by populating the level with a larger orbital quantum number. 

 Axion non-linearities trigger yet another dramatic effect during transitions between Regge trajectories. Namely, attractive axion self-interactions result in a catastrophic instability of the axion cloud and its subsequent collapse. The analogous effect has been observed in laboratory BEC systems with attractive self-interactions and is known under the name ``Bosenova" \cite{Bosenova}. We find that, depending on the parameters, transitions between Regge trajectories proceed through a series of tens to hundreds of Bosenova events. These events produce gravitational waves, and, in the case of the QCD axion, gamma- and X-rays that may be detectable on the Earth.
 
\begin{figure}[t] 
 \begin{center}
 \includegraphics[width=6in]{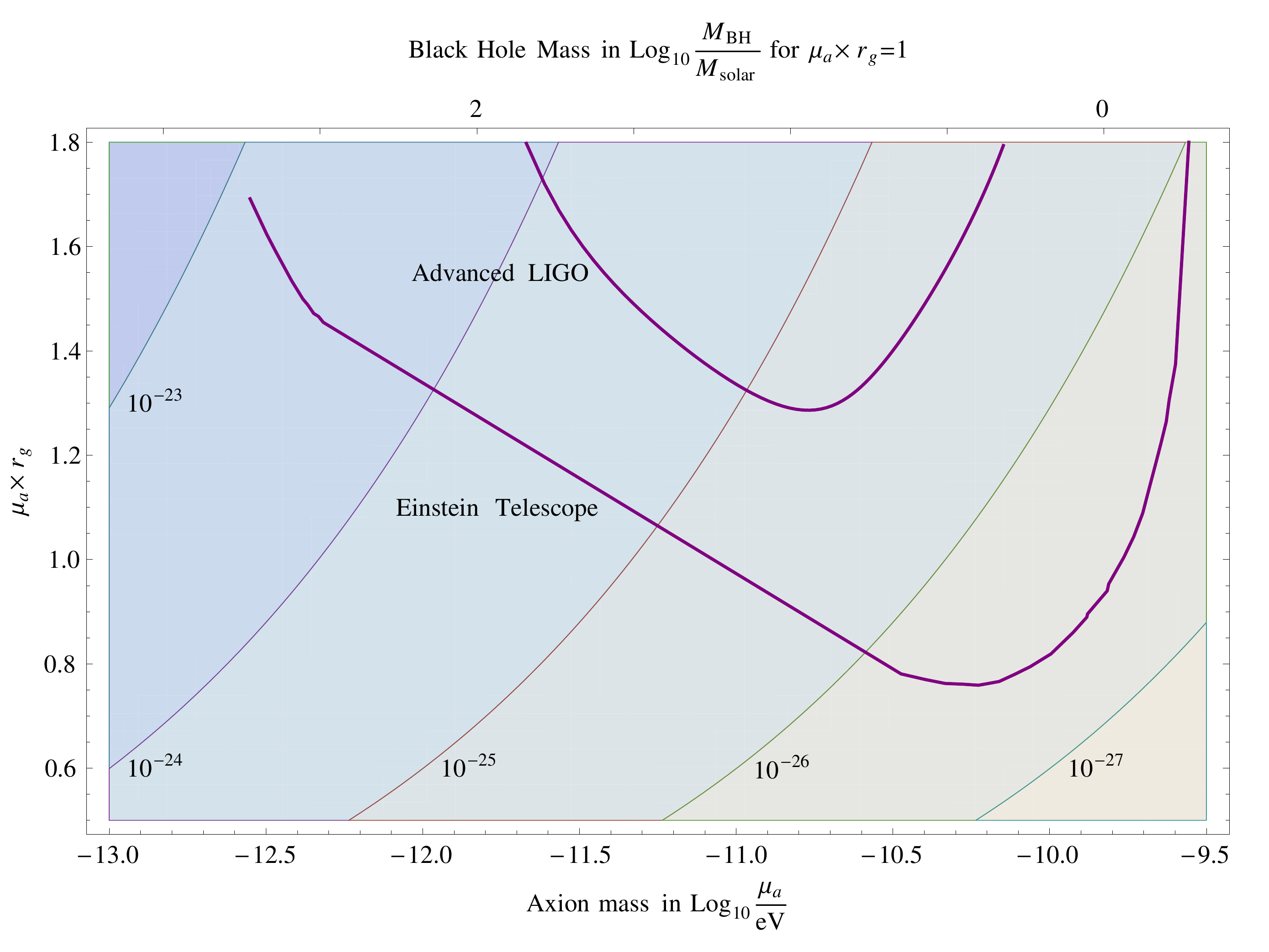}
 \caption{The contour plot of constant gravitational wave signal from axion transitions between the $6g$ and the $5g$ levels for a black hole located at 20~Mpc away from the Earth. The projected sensitivity curves of Advanced LIGO \cite{advLIGO} and Einstein Telescope \cite{ET} assume $10^4$ seconds of a coherent integration time.}
 \label{QCDplot}
 \end{center}
\end{figure}

The black hole may also emit gravitational waves at the observable level due to axion transitions between different levels and annihilations to gravitons. Accurate prediction of the strength and duration of the signal requires a more detailed analysis of the dynamics; in the current paper we limit ourselfs to qualitative estimates which imply that this signal is detectable. It is especially exciting that the gravitational wave signal from the QCD axion cloud around stellar mass black holes falls into the Advanced LIGO frequency band and turns it  into a particle discovery machine.  In Fig.~\ref{QCDplot}  we present the estimated signal strength as a function of the axion and black hole masses for a source at a 20~Mpc distance from the Earth. We choose a coherent integration time for the signal of $10^4$ second , when drawing the sensitivity curves. 
 
The paper is organized as follows. We start with explaining the basics of superradiance in Section~\ref{spectroscopy}.
We argue that the non-relativistic approximation is accurate for describing superradiant levels for most of the parameter space, and adequate at least qualitatively at all values of the parameters.   We present semi-analytical results for the superradiant rates, that are accurate over a large part of the parameter space and enough for our purposes. 

In Section~\ref{dynamics} we discuss the processes that determine the dynamics of the superradiant cloud---gravitational wave emission and axion non-linearities. We provide approximate expressions for the relevant transition rates, and for characteristic timescales and  masses of the superradiant cloud when different processes, such as Bosenova collapse, happen. 

In Section~\ref{signatures} we combine all of the above ingredients and discuss how the superradiant instability develops and what are the associated observational signatures. We discuss in detail the black hole spin-down and of the Regge trajectories. We then estimate the gravitational wave signals during transitions between Regge trajectories and
briefly comment on prospects for directly detecting the axionic cloud around supermassive black holes during extreme mass ratio inspirals. 

We also focus on the potential reach for the QCD axion, and some possible QCD axion specific signatures related to its direct coupling to Standard Model fields. A particularly intriguing effect of the QCD axion is that the effective value of the QCD $\theta$-parameter may become of order one inside the cloud. This possibly destabilizes nuclei in the accretion disk and results in $\gamma$-ray signals and exotic X-ray lines from the nuclear decay products in the black hole vicinity. We finally entertain the possibility of radio waves from axion-to-photon conversion in the near-horizon magnetic field.


Section~\ref{anthropic} elaborates on an issue which is  aside from the main line of the paper, but still a very important part of the theoretical motivations for string axions in the mass range probed by the black hole superradiance. Namely, all these axions are ``anthropic"---their initial misalignment angle needs to be tuned to an atypically small value in the observed part of the Universe. It has been known for a long time~\cite{Linde:1987bx}, that this is not a problem for a single QCD axion, and here we discuss what changes if several anthropic axions are present. 
We conclude in Section~\ref{fuuh}.
 
\section{Spectroscopy of superradiance}
\label{spectroscopy}
In this section we review the spectroscopy of superradiant levels around a rotating black hole. Throughout the paper we are using the Boyer-Lindquist coordinates for the spinning black hole metric \cite{Wald:1984rg}
\bea
&&ds^2 = - (1-\frac{2r_g r}{\Sigma}) dt^2 - \frac{4r_gar \sin^2\theta}{\Sigma} dt d\phi + \frac{\Sigma}{\Delta} dr^2 + 
\Sigma d\theta^2 + \frac{(r^2+a^2)^2 - a^2 \Delta \sin^2 \theta}{\Sigma} \sin^2\theta d\phi^2 \, , \nonumber\\
&&  \Sigma = r^2 + a^2 \cos^2 \theta \, , ~~\Delta = (r-r_+)(r-r_-)~~r_\pm=r_g \pm \sqrt{r_g^2 - a^2} \, , ~~a = \frac{J}{M},~~r_g=G_NM \, , 
\label{kerrsoln}
\eea
 where $M$ and $J$ are black hole's mass and spin, respectively. 
 The physical horizon corresponds to the larger root of $\Delta $, which is $r=r_+$.
 
One of the most fundamental equations for the understanding of superradiance is condition (\ref{Omegacond}), so let us review how it arises, following \cite{Wald:1984rg}. Interestingly, one has to know remarkably little about the Kerr geometry to derive (\ref{Omegacond}). The first fact one needs is that the black hole metric in the Boyer--Lindquist coordinates is invariant under time translations, or more formally it possesses the Killing vector ${\cal H}^\mu\d_\mu=\d_t$. Second, it possesses another Killing vector related to rotational invariance, ${\cal J}^\mu\d_\mu=\d_\varphi$ and the linear combination
\[
{\cal G}={\cal H}+w_+{\cal J},
\]
is normal to the horizon and null there. In the above equation $w_+$ is given by (\ref{spinvelo}). 

Now, let us consider an incoming wave of a scalar field of the form 
\[
\phi=e^{-i\omega t+im\varphi}f(r,\theta)+h.c.\;.
\]
The conserved energy flux of this field is given by
\[
P_\mu=-T_{\mu\nu}{\cal H^\nu}=-\d_\mu\phi\d_t\phi+{1\over 2}g_{\mu t}{\cal L}\;,
\]
where $T_{\mu\nu}$ and ${\cal L}$ are the energy-momentum tensor and the Lagrangian density for $\phi$, respectively.
For a space-time region between two constant time slices, conservation of the current $P_\mu$ implies
that the time-averaged energy flux at the infinity is equal to the time-averaged energy flux through the black hole horizon.
The latter is equal to
\be
\label{horizonflux}
\langle P_\mu{\cal G}^\mu\rangle=-\langle(\d_t\phi+\omega_+\d_\phi\phi)\d_t\phi\rangle=\omega(\omega-m\omega_+)|f|^2\;,
\ee  
where the $g_{\mu\nu}$-term in the energy-momentum tensor dropped out because vectors ${\cal G}$ and ${\cal H}$ are perpendicular at the horizon.
We see that the energy flux is negative in the superradiant frequency range indicating that the wave gets amplified  in this regime.
The argument changes a bit when the frequency $\omega$ corresponds to the discrete spectrum so that the energy flux at the infinity necessarily vanishes. In this case the only way 
to reconcile the flux (\ref{horizonflux}) at the horizon  with the energy conservation is for the frequency $\omega$ to acquire an imaginary part, so that the time-averaged energies on the two constant time slices are not equal any longer. For the real part of the frequency in the superradiant interval (\ref{Omegacond})
the imaginary part should be positive indicating the presence of an instability.

One important consequence of the instability condition (\ref{Omegacond}) is that the superradiant levels are always in a (quasi)non-relativistic Keplerian regime.
Indeed, in this regime the real part of the frequency follows the hydrogen spectrum
\be
\label{hydrogen}
\omega_{\bar{n}}\approx\mu_a\l 1-{\alpha^2\over 2\bar n^2}\r\;,
\ee
where $\bar n=n+l+1$ is the principal quantum number, $l$ is the orbital moment and $\alpha=\mu_ar_g$. For such a level the velocity of the
particle is
\be
\label{hydrogenvel}
v\sim {\alpha\over \bar n}\;.
\ee
On the other hand, if we approximate the frequency in the superradiance condition (\ref{Omegacond}) by the axion mass, $\omega_n\approx\mu_a$, the condition translates into a bound
 \be
 \label{alpha_bound}
 \alpha\lesssim mw_+={m\over 2}{ a\over r_+}\;,
 \ee
 where we made use of  the expression (\ref{spinvelo}) for $w_+$.
  We see now that the velocity for superradiant states may be at most moderately relativistic,
\be
\label{velbound}
v\lesssim \frac{1}{2} {m\over \bar n}{a\over r_+} <{1\over 2}\;,
\ee
where the bound is saturated at the upper boundary of the superradiant range  (\ref{Omegacond}) for extremal black holes $a=r_+$, at $n=0$ and large $l=m\gg1$. 

In principle, this argument does not exclude the presence of a family of non-hydrogenic unstable bound states, however numerical results of \cite{Dolan:2007mj} confirm that all superradiant states are hydrogenic. 

We can also estimate the size of the
axion cloud as
\be
\label{cloudsize}
r_c\sim {\bar n^2\over \alpha^2}r_g\;,
\ee
which is always significantly larger than the black hole gravitational radius, as a consequence of (\ref{Omegacond}). 

These estimates provide the following physical picture of the superradiant axion cloud. The cloud is composed of a wave packet of the axion field rotating on a Keplerian orbit around the black hole. This axion wave packet always has a tail that goes into the near-horizon ergosphere region and gets amplified there leading to the exponential growth of the number of axions in the packet.
 
 Given the complexity of the Kerr metric it is not surprising that a precise analytical expression for superradiant rates is unavailable (partial numerical results can be found in \cite{Dolan:2007mj}). However, the above physical picture gives rise to two useful analytical approximations for the superradiant rates. Before introducing them let us recall that the massive Klein--Gordon equation in the Kerr background allows separation of variables \cite{Brill:1972xj} with the following simple ansatz for the scalar field
 \[
\phi=e^{-i\omega t+im\varphi}Y(\theta)R(r)+h.c.\;.
\]
The equation for $Y(\theta)$ is the standard equation for the flat space spherical harmonics plus an extra term that can be neglected if $\l\alpha/l\r^2\ll 1$. 
As before, the superradiant condition (\ref{Omegacond}) implies that $\l\alpha/l\r^2< 1/4$, so we will always use this approximation (one can check its accuracy using known numerical results for oblate spheroidal harmonics, see e.g. \cite{Berti:2005gp}). Then the equation for the radial function $R$ takes the form 
\be
\label{Req}
\Delta\d_r(\Delta \d_r R)+\l \omega^2(r^2+a^2)^2-4ar_grm\omega+a^2m^2-\Delta(\mu_a^2r^2+a^2\omega^2+l(l+1)\r R=0\;.
\ee
At the horizon, a non-singular solution of this equation satisfies \cite{Press:1973zz}
\be
\label{inwave}
R=const \cdot e^{-i (\omega-mw_+)r_*}\;\;\mbox{as}\; r\to r_+\;,
\ee
where $r_*$ is the ``tortoise" coordinate defined through
\be
\label{tortoise}
dr_*=(r^2+a^2)\Delta^{-1}dr\;.
\ee
 
 \subsection{Non-relativistic approximation $\alpha/l\ll1$}
 \label{lowalpha}
 This approximation \cite{Ternov:1978gq,Detweiler:1980uk}, that initially was applied for superradiant scattering rather than the calculation of the instability rate \cite{Starobinskii}, makes use of the separation of scales between the size of the cloud and the black hole horizon following from relations (\ref{alpha_bound}), (\ref{cloudsize}). The radial equation is now solved in two different regimes: the near and far horizon regions. In the region far from the black hole horizon, $r\gg r_g$, neglecting terms suppressed by $\l{\alpha/ l}\r^2$, the solution takes the same form as the radial wave function of the Schroedinger equation with an $1/r$ potential,
 \be
 \label{Rfar}
 R_{far}(r)=(2kr)^le^{-kr}U(l+1-{\alpha^2\over r_g k},2(l+1), 2kr)\;,
 \ee
 where $U$ is the confluent hypergeometric function of the second kind, and $k$ is the axion momentum,
 \be
 \label{kdef}
 k^2=\mu_a^2-\omega^2\;.
 \ee
 In the ordinary Schroedinger equation with an $1/r$ potential the spectrum of frequencies $\omega$ is determined by requiring the regularity of $R$ at the origin. Instead, in the black hole case one has to impose the regularity of the field at the horizon---the incoming wave  boundary condition (\ref{inwave}). One way to do this is to solve the radial equation (\ref{Req}) in the near horizon regime. After dropping  terms of order $\alpha/l$ the solution in the near-horizon region, $0<r-r_+\ll (l/\alpha)^2r_g$,  that satisfies the boundary condition (\ref{inwave}) takes the form
 \be
 \label{Rnear}
 R_{near}(r)=\l{r-r_+\over r-r_-}\r^{-i P} \;_2F_1(-l,l+1,1+2iP,{r-r_-\over r_+-r_-})\;,
 \ee
 where 
 \[
 P=2r_+{\omega-mw_+\over r_+-r_-}
 \]
  and $_2F_1$ is the Gauss's hypergeometric function. 
  At $\alpha/l\ll1$ the ranges of validity for  the two approximate solutions (\ref{Rfar}) and (\ref{Rnear}) have an overlap, so the approach in \cite{Ternov:1978gq,Detweiler:1980uk} was to match the lowest terms of the Taylor expansion for $R_{far}$ at small $r$ with the asymptotic behavior of (\ref{Rnear}) at large $r$, with the following result for the imaginary part of the frequency,
   \begin{gather}
   \label{Detweiller}
   \Gamma_{lmn}=2\mu\alpha^{4l+4}r_+(mw_+-\mu_a)C_{lmn}\;,
  \end{gather}
where
\[
C_{lmn}= 
  {2^{4l+2}(2l+n+1)!\over (l+n+1)^{2l+4}n!}\l{l!\over(2l)!(2l+1)!}\r^2
    \prod_{j=1}^l
   \l j^2\l1-{a^2\over r_g^2}\r+4r_+^2(mw_+-\mu_a)^2\r\;.
   \]
    Note that the real part is well approximated by the hydrogen spectrum (\ref{hydrogen}). These approximate formulaes exhibit many of the features of the full answer. In particular, the sign of $\Gamma_{lmn}$ is determined by the sign of $(mw_+-\mu_a)$, in agreement with (\ref{Omegacond}) within the accuracy of the non-relativistic approximation used to derive (\ref{Detweiller}). Also, in the regime of applicability of (\ref{Detweiller}), $\alpha/l\ll1$, widths $\Gamma_{lmn}$ drop exponentially as $l$ increases. This implies that for a given value of $\alpha$ the fastest superradiant level is the one with a smallest possible $l$, {\it i.e.}, the  $l=m$ level with $m$ chosen in such a way that the superradiance condition (\ref{Omegacond}) is satisfied.
   
   It will be important in what follows that as $l$ grows the radial quantum number $n$ for the fastest superradiant level grows as well. For instance, using (\ref{Detweiller}) we find that for $l=m=4$ the instability rate for the $n=1$ level is faster than for the $n=0$ level,
   \be
   \label{4410}
   {\Gamma_{440}\over \Gamma_{441}}\approx 0.9.
   \ee

   The result (\ref{Detweiller}) is often referred to as  the ``low mass" or small $\alpha$  approximation. However, as the above derivation shows, the approximate solutions (\ref{Rfar}) and (\ref{Rnear}) hold and have an overlapping regime of validity at large $\alpha$ as well, as long as $\alpha/l\ll 1$. It is true, though, that  (\ref{Detweiller})  may be  quite inaccurate close to the superradiance boundary $\alpha\lesssim w_+$, both due to the inaccuracy of the approximate solutions and because the overlap interval where both of the solutions hold shrinks.
   
\begin{figure}[t] 
 \begin{center}
 \includegraphics[width=7in,trim= 0 70 0 70]{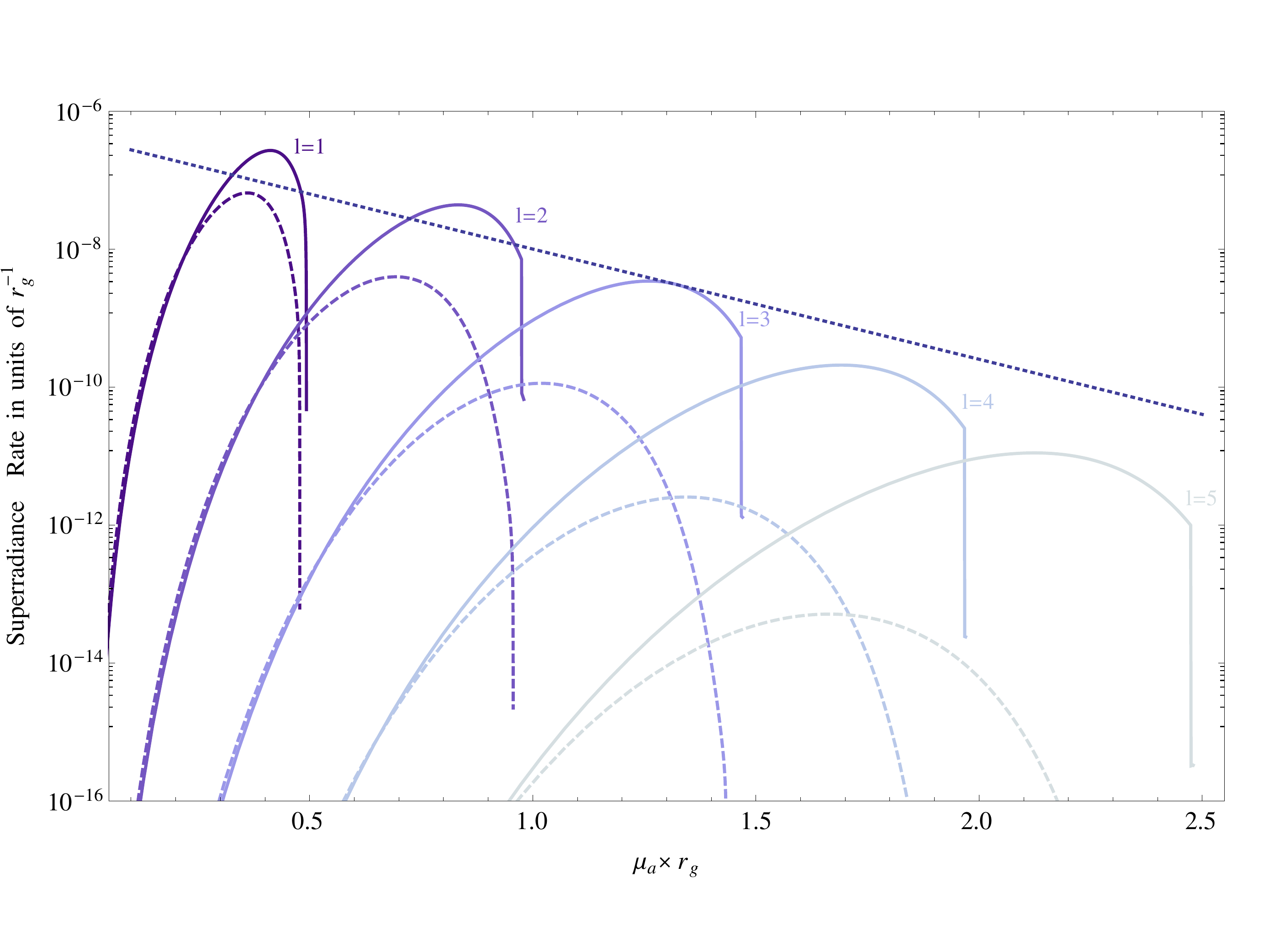}
 \caption{Superradiance rates obtained using our semi-analytic method (solid lines), non-relativistic approximation (dashed lines) and WKB approximation (dotted line) for a near-extremal black hole, $a/r_g=0.999$.
 Different colors correspond to superradiant levels with different values of the angular quantum number $l$.}
 \label{rates}
 \end{center}
\end{figure}

 To have better control of the precision and to improve on the latter defficiency of (\ref{Detweiller}) we adopt the following semi-analytic procedure, which is similar to that in \cite{Rosa:2009ei}. Instead of matching the leading terms in the asymptotic expansions of (\ref{Rfar}), (\ref{Rnear}) we numerically matched these functions and their first derivatives at a point $r_*$ within the overlap region. To find the optimal value for $r_*$ we calculated the relative residuals after plugging $R_{far}$ and $R_{near}$ in the original radial equation (\ref{Req}). We pick $r_*$ as the point where the two residuals are equal. We present our semianalytical results for the instability rates in Fig.~\ref{rates} (solid lines) together with the instability rates given by (\ref{Detweiller}) (dashed lines). Different lines correspond to different $l=m$ levels, and  we picked the radial quantum number $n$
  to maximize the superradiance level in each case  (the dependence on $n$ is very mild). 

\begin{figure}[t] 
 \begin{center}
 \includegraphics[width=7in,trim= 0 70 0 70]{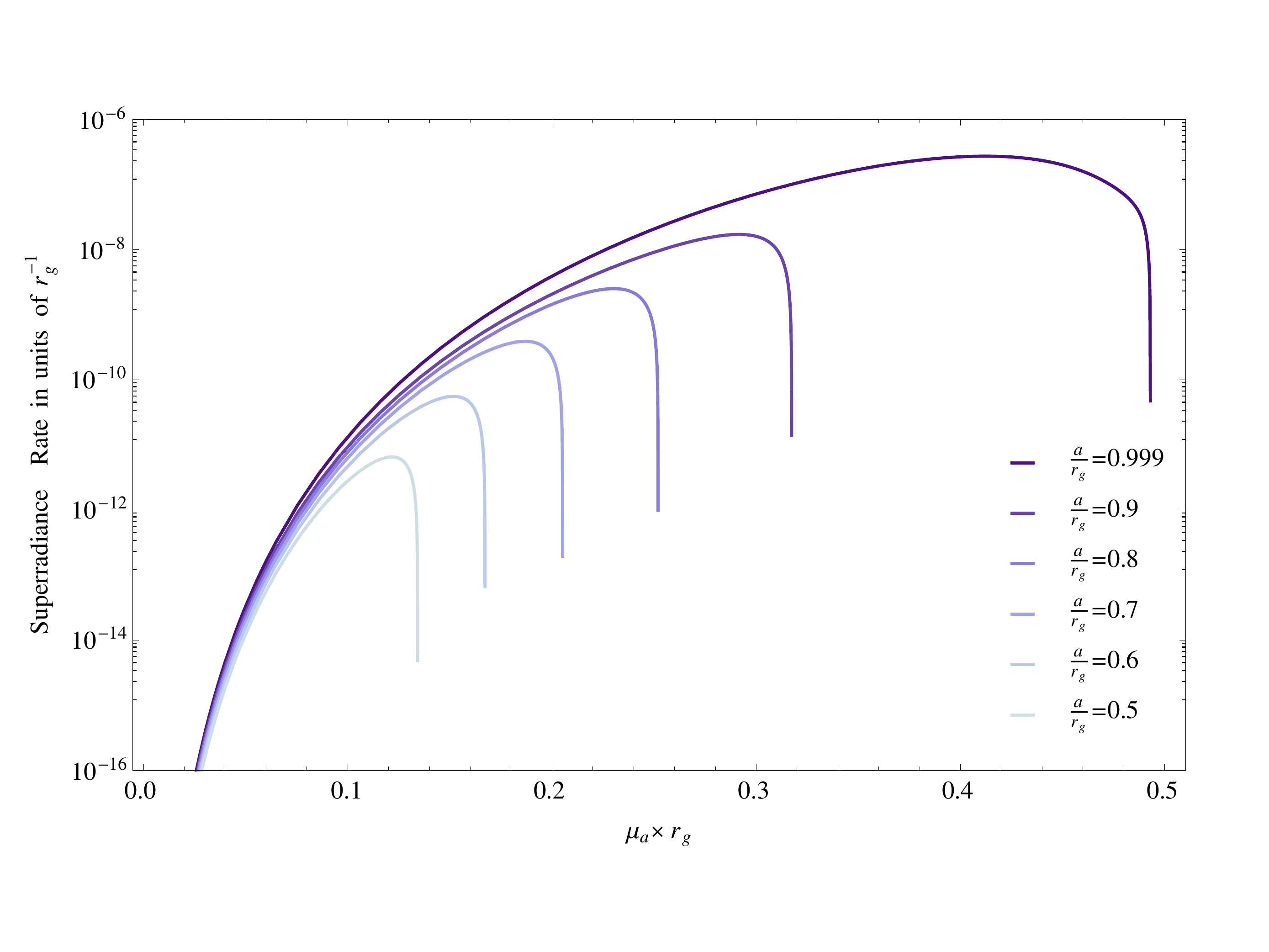}
 \caption{Superradiance rates obtained using our semi-analytic method (solid lines) for different values of the black hole spin.}
 \label{l1rates}
 \end{center}
\end{figure}

  These results were obtained for a near extremal black hole with $a/r_g=0.999$. The instability rate decreases very slowly with spin for small $\alpha$, and the main effect of reducing the spin is that the instability shuts down earlier, at $\alpha_m=mw_+(a)$
  for the different $l=m$ levels, as follows from the superradiant condition (\ref{Omegacond}). To illustrate this in Fig.~\ref{l1rates} we present the superradiance rates for $l=1$ level for several values of $a/r_g$.

 We see that (\ref{Detweiller}) perfectly agrees with our semi-analytic results at sufficiently small $\alpha/l$, but they quickly start being different indicating that the procedure of matching the leading terms of the asymptotic expansions that results in (\ref{Detweiller}) is not very accurate. Of course, our semi-analytic results are also not good enough for precision calculations close to the superradiant boundary, $\alpha\sim mw_+$, however, they agree quite well with numerical calculations in  \cite{Dolan:2007mj} (at least for $l=1,2,3$ presented in  \cite{Dolan:2007mj} ) and with the WKB results presented in the next subsection. In particular, the superradiance rate of  \cite{Dolan:2007mj} is maximum, $\Gamma\approx 1.5\cdot 10^{-7}r_g^{-1}$, at $\alpha\approx 0.42$ in a good agreement with our results. As we will see, many of the observational consequences of superradiance are not very sensitive to the exact values of the superradiance rates at $\alpha\sim mw_+$, so this level of precision is enough for our purposes.
  \subsection{WKB approximation $\alpha\gg 1$}
 \begin{figure}[t!] 
 \begin{center}
 \includegraphics[width=7in,trim=0 150 0 150]{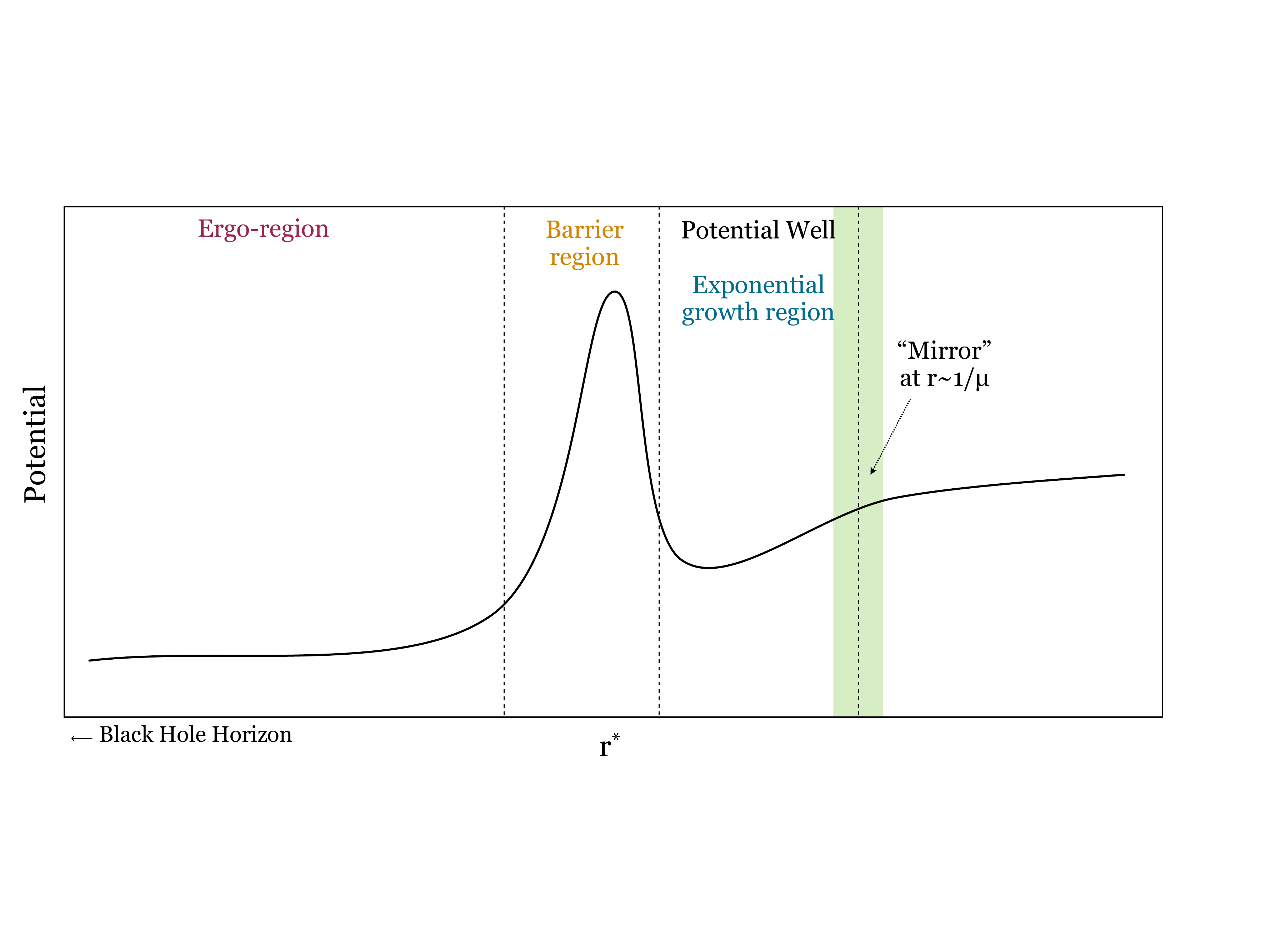}
 \caption{The shape of the radial Schroedinger potential for the eigenvalue problem in the rotating black hole background. Superradiant modes are localized in a potential well region created by the mass ``mirror" from the spatial infinity on the right, and by the centrifugal barrier from the ergo-region and horizon on the left.}
 \label{fig:potential}
 \end{center}
\end{figure}
Another useful approximation for the superradiant rates \cite{Zouros:1979iw}, complementary to the slow velocity expansion above, is the WKB method that can be applied at $\alpha\gg1$. We will closely follow the methodology of  \cite{Zouros:1979iw}, however, our results disagree  with  \cite{Zouros:1979iw} by an important factor of two in the expression for the tunneling exponent. Most likely, this factor was accidentally missed in  \cite{Zouros:1979iw} (this discrepancy was also pointed out in  \cite{Gaina:1989sj} without any derivation).
 
 In this approach, the tail of the wave function that propagates in the ergo-region, where superradiant amplification takes place,
is calculated using the WKB approximation. This is just a classic tunneling calculation. Indeed,  after switching to the tortoise coordinate (\ref{tortoise}) and introducing $\Psi=(r^2+a^2)^{1/2} R$ the radial equation (\ref{Req}) takes the form of the Schr\"odinger equation
 \be 
 \label{Shroed}
 {d^2\Psi\over dr_*^2}-V\Psi=0 
 \ee
 with the potential 
 \be
 \label{potential}
 V=  -\omega^2+{4 r_g ram\omega-a^2m^2\over (r^2+a^2)^2}+{\Delta\over r^2+a^2}\l {\mu_a^2}+{ l(l+1)+k^2 a^2\over r^2+a^2}+{3 r^2-4 r_g r+a^2\over (r^2+a^2)^2}-
 {3\Delta r^2\over (r^2+a^2)^3}\r
 \ee
 We include the $(-\omega^2)$ term in the definition of the potential, because even if we were to separate it, there would be a residual dependence on $\omega$.
We present  the qualitative shape of the potential  $V$ for a typical choice of parameters in Fig.~\ref{fig:potential}. One can clearly see the potential well where the bound Keplerian orbits are localized and a barrier separating this region from the near-horizon region where superradiant amplification takes place. 
 
 Consequently, the axion wave function at the horizon $r=r_+$ (corresponding to $r_*=-\infty$)
is suppressed relative to the wave function in the vicinity of the Keplerian orbit by a tunneling exponent,
\[
|R(r_+)|\simeq|R(r_c)|e^{-I}\;,
\]
 where the tunneling integral $I$ is
 \be
 \label{Integral}
 I=\int_{r_*(r_1)}^{r_*(r_2)}dr_*\sqrt{V}=\int_{r_1}^{r_2}dr{\sqrt{V}(r^2+a^2)\over\Delta}\;,
 \ee
 with $r_{1,2}$ being the boundaries of the clasically forbidden region.  We will only follow the leading exponential dependence on $e^{-I}$ and do not aim at calculating the normalization prefactor in front of the exponent.
 
To relate the tunneling exponent with the rate of superradiance instability let us consider again the energy flow equation (\ref{horizonflux}). Integrating it over the horizon we obtain
\be
\label{Edot}
{d{\cal E}\over dt}=\omega(mw_+-\omega)\int_{horizon}|Y(\theta)R(r_+)|^2\;,
\ee
where ${\cal E}$ is the energy in the axion cloud. The energy is maximum in the Keplerian region, so that in the limit where we only keep track of the dependence on the exponent $e^{-I}$ we can write
\[
{\cal E}\propto |R(r_c)|^2\simeq e^{2I} |R(r_+)|^2\;,
\] 
and, consequently, to rewrite (\ref{Edot}) as
\be
{d{\cal E}\over dt}=const\cdot (mw_+-\omega)e^{-2I}{\cal E}\;.
\ee
 In other words, the WKB approximation for the superradiance rate gives\footnote{Note, that at this stage we still agree with \cite{Zouros:1979iw}.}
 \be
 \label{GammaWKB}
\Gamma=\gamma(mw_+-\omega)e^{-2I}\;,
 \ee
 where the normalization prefactor is determined mainly by the spread of the wave function in the classically allowed region. We will limit ourself by calculating the exponential part $\Gamma$. 
  We leave the technical details for the Appendix, and present only the final result here.
  Namely,  the final answer for the tunneling integral in the extremal Kerr geometry takes the  form
 \be
 \label{finalI}
 I=\pi \l 2\alpha-\sqrt{2\alpha(\alpha-1)}\r\;,
 \ee
 which translates in the following superradiant rate,
 \be
 \label{WKBrate}
 \Gamma_{WKB}\approx 10^{-7}r_g^{-1}e^{-2\pi\alpha(2-\sqrt{2})}\approx 10^{-7}r_g^{-1}e^{-3.7\alpha}\;,
 \ee
 where we took the large $\alpha$ limit in (\ref{finalI}) and chose the prefactor to match the low $\alpha$ results of section~\ref{lowalpha} (this value also agrees with that of \cite{Zouros:1979iw} and \cite{Gaina:1989sj}). As we already said, the exponent in (\ref{WKBrate}) is larger than that in
 \cite{Zouros:1979iw} by a factor of two. As explained in the Appendix, the rate (\ref{WKBrate}) provides an upper envelope for superradiance rates at different $l$ in
 the large $\alpha$ limit.
 We have presented (\ref{WKBrate}) by a dotted line in Fig.~\ref{rates}; it agrees reasonably well  with the previous $\alpha/l\ll 1$
 results.

\section{Dynamics of superradiance}
\label{dynamics}
Let us turn now to discussing the dynamical consequences of the superradiant instability. One important property of the rates calculated in section~\ref{spectroscopy} is that the time-scale for the development of the instability  is quite slow compared to the natural dynamical scale $r_g$ close to the black hole horizon, $\Gamma_{sr}^{-1}>10^{7}r_g$. Consequently, in many cases  non-linear effects, both gravitational, and due to axion self-interactions, become important in the regime where the system is still in the quasi-linear regime, so that non-linearities can be treated perturbatively.  Then the dynamics of the axion cloud can be described by the following set of kinetic equations for the occupation numbers $N_i$ for different levels,
\be
\label{kineq}
{dN_i\over d t}=\Gamma_{ij}N_j+\Gamma_{ijk}N_jN_k+\dots\;. 
\ee
This equation gets simplified in the quasilinear regime where we can truncate the expansion in the r.h.s. of (\ref{kineq}), by keeping just a finite number of terms. Note that for simplicity, we drop $N_i$-independent terms in the r.h.s. of (\ref{kineq}); those terms correspond to spontaneous emission.

To avoid confusion, let us clarify the following. Throughout this paper we often use  quantum terminology (occupation numbers, transition between levels, etc.) to describe the axionic cloud. This appears to be perfectly appropriate given that the size of the cloud is comparable to the Compton wavelength of the axion. On the other hand, occupation numbers for all dynamically relevant levels will always be exponentially large in what follows, $\sim 10^{70}$, so that all the dynamics can be accurately described by a classical field theory.  Of course, there is no contradiction here and both descriptions are correct. The very fact that we can use classical {\it field} theory to describe the dynamics of axion {\it particles} in the cloud reflects its quantum mechanical origin. For instance, in the classical field description instead of using occupation numbers $N_i$  one can Fourier decompose  classical field into harmonics with different frequencies and follow the (squared) amplitudes for different harmonics.
Given numerous analogies with atomic physics we find the quantum language useful in many cases, but will also use the classical one, when more convenient. 

Coming back to the kinetic equations (\ref{kineq}), 
at the linear level, the r.h.s. of (\ref{kineq}) is determined by the superradiant rates presented in section~\ref{spectroscopy}.
Namely, 
\be
\label{linG}
\Gamma_{ij}=\delta_{ij}\Gamma_{i}\;,
\ee
where $\Gamma_{i}$ is the imaginary part of the frequency for the $i$-th level---positive for levels satisfying (\ref{Omegacond}) and negative otherwise.
There could be other model dependent sources of linear terms in (\ref{kineq}). For instance, if an axion has an electromagnetic coupling 
\be
\label{axphot}
{C \alpha\over 4\pi f_a}\phi\epsilon^{\mu\nu\lambda\rho}F_{\mu\nu}F_{\lambda\rho}\;,
\ee
where $C$ is an order one constant (for the QCD axion $C=4/3$ in 4d grand unified theories),
then in the presence of a magnetic field axions will convert into photons 
with a rate \cite{Arvanitaki:2009fg}
\be
\label{agamma}
\Gamma\sim7\cdot 10^{-11}\mbox{yr}^{-1} \l{10^{16}\mbox{GeV}\over f_a}\r^2\l{\mu_a\over 6\cdot 10^{-10}\mbox{eV}}\r\l {B\over 4\cdot 10^8 \mbox{G}}\r^2\;,
\ee
where the reference values for the parameters are chosen to be those for the QCD axion, and the choice of a reference magnetic field is motivated by the estimate
(Eq.~(75) of \cite{Rees:1984si})
\be
B\sim 4 \times 10^8 G \l{M\over M_\odot}\r^{-1/2} \, 
\label{Bbh}
\ee
for the largest magnetic field the accretion disc can support near the horizon of a black hole of mass $M$. It is evident from (\ref{agamma}) that axion-photon conversion is too slow to be relevant for the dynamics of superradiance. 

To describe the development of the superradiant instability one needs to supplement (\ref{kineq}) with equations for the time evolution of the black hole mass $M$ and spin $J$. These depend on the environment of each individual black hole, but, in general, accretion in the absence of mergers or other violent events has a characteristic time scale whose lower bound is set by the Eddington time $\tau_E$ 
\be
\label{Eddington}
\tau_E\equiv {\sigma_T\over 4\pi G_Nm_p}\approx 4\cdot10^8 \mbox{~yr}\;.
\ee
The value of $\tau_E$ shows that the superradiant instability time, even though much shorter than the black hole infall time, is much faster than the evolution time scale for astrophysical black holes. As a result, for most of the discussion that follows we will ignore accretion, unless stated otherwise. For example, in section \ref{bhgrowth}, we discuss a particular accretion model. 

Let us describe now the leading non-linear processes, which should be included in (\ref{kineq}) to describe the development of the superradiant instability.
The two sources of non-linearities are gravitational interactions of axions and non-linearities in the axion potential itself.

\subsection{Gravity wave emission}
\label{gravemission}

The axion cloud may lose its energy and angular momentum by emitting gravitational waves. There are two major processes giving rise to graviton emission. The first is analogous to photon emission from atoms --- this is just the transition of axions from one level to another. The major difference in the present case is that the transition rates between populated levels get enhanced by their occupation numbers. 

The other process of graviton emission is less familiar. Unlike electrons in the atom, axions do not carry any conserved charge. Consequently, they can emit gravitational waves also through processes that do not conserve the axion number. In particular, one-graviton annihilation of two axions is possible at the same order of perturbation theory as the transition between different levels, {\it i.e.}, the corresponding amplitude is proportional to $M_{Pl}^{-1}$. Of course, this process is kinematically forbidden in flat space, where only two-graviton
annihilation (with amplitude proportional to $M_{Pl}^{-2}$) is compatible with energy and momentum conservation. However, the black hole  gravitational field breaks translational invariance, so that one-graviton annihilation is allowed for axions in the presence of a black hole. The closest analogue of this process in atomic physics is the one-photon annihilation of a positron with an atomic electron \cite{Fermi}.

This annihilation process is quite unconventional, because the graviton 
momentum is determined by the axion mass $k_g\approx 2\mu_a$ and is parametrically larger than the momentum of axions in the cloud $k_a\sim (\alpha/l)\mu_a$. In other words, unlike for conventional astrophysical sources of gravitational waves, the wavelength of an emitted radiation is {\it not}
parametrically longer  than the size of the source (even though the cloud is non-relativistic), and the standard quadrupole approximation does not apply.

\subsubsection{General formalism}
The calculation of the gravitational wave flux due to both kind of processes is further complicated by the fact that the system is bound by gravity, and the
axion cloud is not that far from the black hole horizon, where the gravitational field is non-linear. However, these complications can be safely ignored for order of magnitude estimates, which are enough for our purposes. To perform the estimates it is convenient to switch to the classical field theory description. Let us write the axion field of the cloud in the form
\be
\label{field_harmonics}
\phi=\sum_\omega e^{-i\omega t}\chi(\omega,\varphi,r,\theta)+h.c.\;,
\ee
where the frequencies $\omega$ ran over different bound levels in the black hole geometry. The related expansion of the scalar energy-momentum tensor takes the form
\be
\label{Tharmonics}
T_{\mu\nu}=\sum_{\omega'} e^{-i\omega' t}\tau_{\mu\nu}(\omega',\varphi,r,\theta)+h.c.\;.
\ee
To estimate the gravitational wave flux from the axion cloud we use the flat space formula \cite{Weinberg} for the gravitational wave power,
\be
\label{weinberg_power}
{dP\over\sin\theta d\theta d\varphi} ={G_N\omega^2\over \pi}\tau_{ij}^{TT*}(\omega,k)\tau_{ij}^{TT}(\omega,k)\;,
\ee
where 
\be
\label{fourrier}
\tau_{ij}(\omega,k)\equiv\int d^3{\bf{x}}\tau_{ij}(\omega,{\bf{x}})e^{-ik{\bf x}}
\ee
where ${\bf{x}}$ denotes the flat space Descartes coordinates, $|k|=\omega$, and the $TT$ superscript stands for the projector on the transverse-traceless part,
\[
\tau_{ij}^{TT}\equiv ( P_{ii'}P_{jj'}-{1\over 2} P_{ij}P_{i'j'})\tau_{i'j'}^{TT}\;,
\]
where 
\[
P_{ij}=\delta_{ij}-{k_ik_j\over k^2}\;.
\]
In this classical language the graviton emission due to axion transitions between levels correspond to terms in the harmonic expansion
(\ref{Tharmonics}) of the energy-momentum tensor of the form\[
e^{-i(\omega-\omega')t}\chi(\omega)\chi^*(\omega')+h.c.\;,
\]
where,  for simplicity, we dropped all the derivatives appearing in the expression for
$T_{\mu\nu}$.
On the other hand, the annihilation processes correspond to terms of the form
\[
e^{-i(\omega+\omega')t}\chi(\omega)\chi(\omega')+h.c.
\]
To calculate the gravitational wave power, one needs to solve for the scalar field harmonics $\chi(\omega)$ in the black hole background and plug them
into (\ref{weinberg_power}). This is similar to calculating the trajectory of a compact stellar mass object falling into a galactic mass black hole in the probe particle approximation, and using this trajectory as a source in the linearized Einstein equations to solve for the $h_{ij}$ components of the metric, that determine the flux of gravitational waves.
In principle, the latter step should be done in the curved geometry---the flat space approximation is just an estimate
both for the overall rate and for the frequency profile of the gravitational wave signal. 
For instance, for the annihilation signal from a single populated level with the frequency $\omega$ the flat space expression (\ref{weinberg_power})  predicts the monochromatic gravitational wave line of frequency $2\omega$, while taking into account 
deviations from the flat space would induce the broadening of the line due to gravitational redshift. 

However, given the separation of scales between  the size of the axion cloud and the black hole size, we expect that by using the flat space expression (\ref{weinberg_power})  we are making at most order one 
mistake in the overall rate, which is accurate enough for our purposes. For the same reason, the spectral distortions should be rather small. It is likely that one still needs to calculate more accurately the spectral shape of the signal (the waveform) to be able to observe it with gravitational wave detectors. This is a technically involved calculation, which is beyond the scope of the current paper. Note that in 
\cite{Babak:2006uv}, where gravitational waveforms in the Kerr metric are calculated, the flat space approximation gives quite accurate results.

The other approximation we adopt to evaluate (\ref{weinberg_power}) is rather than solving for the eigenfunction $\chi(\omega,\varphi,r,\theta)$ in the full Kerr background we will just use the Newtonian approximation for the metric. Again, this is justified because the cloud is mainly localized relatively far from the black hole horizon. In this approximation the eigenfunctions $\chi(\omega,\varphi,r,\theta)$  are the familiar wave functions for the electron states in a hydrogen atom, and the evaluation of (\ref{weinberg_power})  becomes straightforward. Let us present here some representative results of this calculation that we will use later in section~\ref{signatures}.

\subsubsection{Axion transitions between levels in a black hole atom}
Let us start with the more familiar case of graviton emission due to axion transitions between levels. As we will see in section~\ref{signatures} an important 
source of potentially observable gravitational radiation is related to axion transitions between levels with equal angular quantum numbers $l$ and $m$, but with different radial quantum numbers $n$. This process will be relevant for sufficiently high $l$, such that the fastest superradiant level has $n>0$. For instance, 
 as we saw in section~\ref{lowalpha}, for $l=m=4$ level the $n=1$ superradiant level is faster than $n=0$, Eq.(\ref{4410}). 
 For the transition between these two levels (\ref{weinberg_power}) gives
 \be
 \label{gpower}
 {dP\over\sin\theta d\theta d\varphi}(6g\to 5g)\approx N_1N_0 {2^{23} 3^410^5 G_N \alpha^{12}\over 11^{22}\pi r_g^4}\sin^4\theta +\dots\approx 3\cdot 10^{-10}N_1N_0{G_N\alpha^{12}\over r_g^4}\sin^4\theta\;,
 \ee
where $N_1$ and $N_0$ are the occupation numbers for the two levels and dots stand for higher order terms in the small $\alpha$ expansion (we checked that they can be neglected in the superradiant regime $\alpha\lesssim 2$). 

For transitions between levels the wavelength of an emitted graviton is much longer than the size of the system, so the conventional quadrupole formulae for a graviton emission should also be a good approximation. As a cross-check of our calculation let us compare result (\ref{gpower}) with the quadrupole formula.
For the transition rate between two levels the latter gives \cite{Weinberg}
\be
\label{Wrate}
\left.\frac{dN_1}{dt}\right|_{quadr}=N_1N_0{2 G_N\Delta \omega^5\over 5}I_{ij}I_{ij}\;,
\ee
where $\Delta \omega$ is the frequency splitting, which in our case is equal to
\be
\label{deltaw}
\Delta \omega={\mu_a\alpha^2\over 2}\l{1\over  25}-{1\over 36}\; \r,
\ee
and $I_{ij}$ is the transition mass quadrupole moment. For the estimate we take
\[
I_{ij}I_{ij}\sim\mu_a^2r_c^4\;,
\]
where the size of the cloud $r_c$ is estimated by (\ref{cloudsize}). Altogether, this gives
\be
\label{gqrate}
\left. \frac{dN_1}{dt}\right|_{quadr}(6g\to 5g)\sim 8\cdot 10^{-7}N_1N_0{G_N \alpha^{9}\over  r_g^3}\;.
\ee
On the other hand, from (\ref{gpower}) we get
\be
\label{grate}
 \frac{dN_1}{dt}(6g\to 5g)={\int_{angles} dP\over\Delta w}\approx 3\cdot 10^{-7}N_1N_0{G_N \alpha^{9}\over  r_g^3}
\ee
in a perfect agreement with (\ref{gqrate}).

Proceeding as above it is straightforward to calculate other transition rates. For instance, for transitions from the fastest ($n=0$) $l=2$ superradiant level to the fastest (also $n=0$) $l=1$ superradiant level one gets (for simplicity, we present only the total power integrated over directions)
\be
\label{dprate}
 {P}(3d\to 2p)\approx N_1N_0{5717\cdot 2^8 G_N\alpha^{14}\over 3^5 5^{11}7^3r_g^4}\approx 4\cdot 10^{-7}N_1N_0{G_N\alpha^{14}\over r_g^4} \;.
\ee
This rate is suppressed by a higher power of $\alpha$, because the quadrupole transition between these two levels is forbidden.

\begin{figure}[t] 
 \begin{center}
 \includegraphics[width=7in,trim=00 60 0 0]{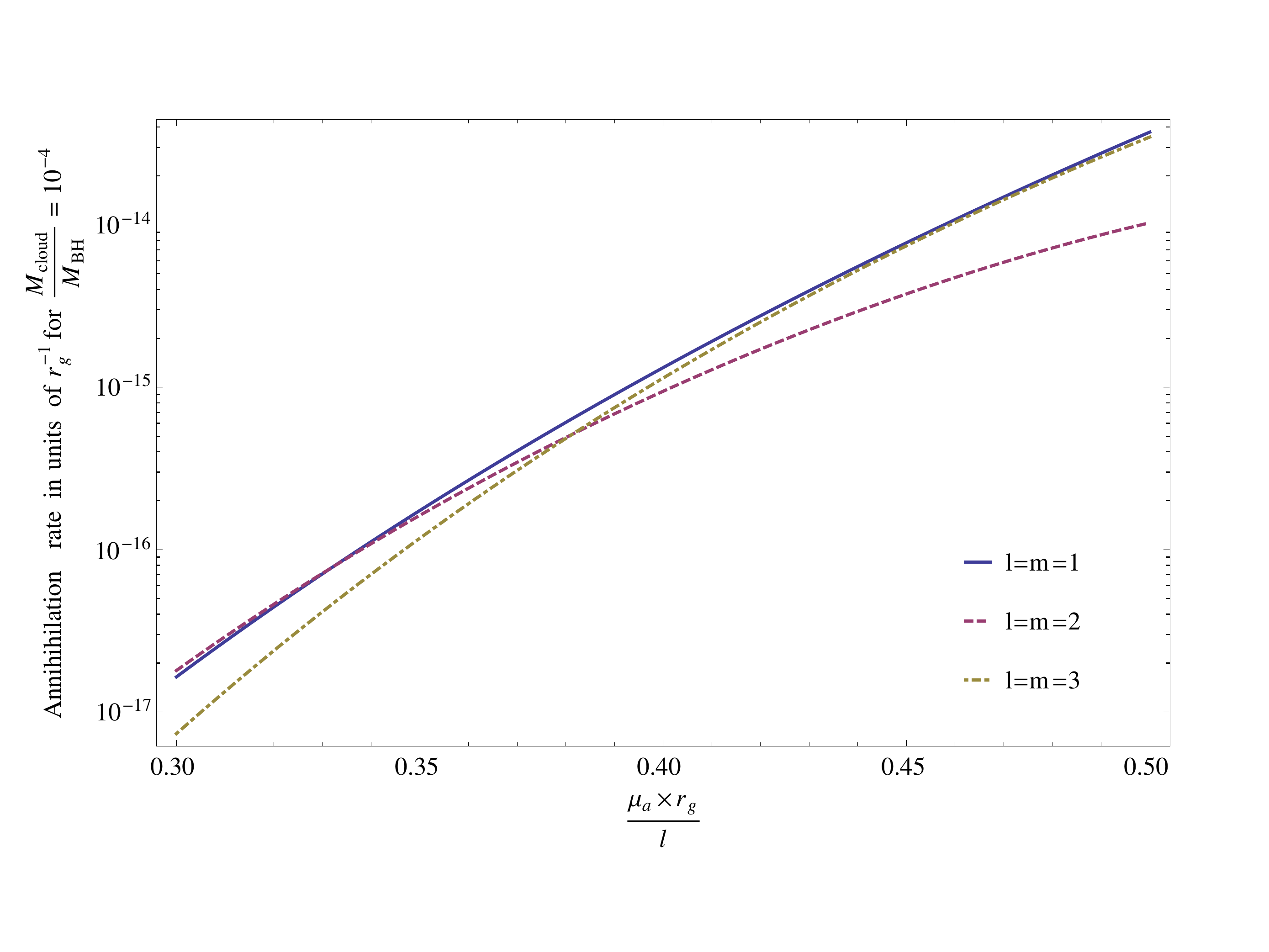}
 \caption{Total annihilation rates in units of $r_g^{-1}$ for $\frac{M_{cloud}}{M_{BH}}=10^{-4}$ as a function of $\alpha/l$ for the superradiant levels with $l=1,~2,~3$.}
 \label{annihilation}
 \end{center}
\end{figure}
\subsubsection{Axion annihilations}
We see that for estimating the transition rates one can use  the standard multipole formulae.
As we said, this is no longer the case for annihilations, where the wavelength of an emitted graviton is shorter than the size of the cloud.
In this case the suppression for the emission rate is related to the decoupling of high momentum modes---the Fourier transform in (\ref{fourrier}) involves convolution of the slowly 
varying energy-momentum tensor of the axion cloud with a rapidly oscillating exponent. A direct calculation gives the following result for the annihilation rate
at the $2p$ level, the fastest superradiant level,
\be
\label{11annihilation}
{dP\over\sin\theta d\theta d\varphi}(2\times 2p\to graviton)\approx N^2 {9 \pi G_N \alpha^{18}\over 2^{26}r_g^4}(35+28\cos{2\theta}+\cos{4\theta})\;,
\ee
where we again expanded the full answer at small $\alpha$ and $N$ is the occupation number. Note that, unlike for transitions, the corrections from higher order terms in $\alpha$ 
change the answer for the annihilation rate by a factor of order one close to the upper boundary of the superradiant regime (\ref{Omegacond}) ($\alpha\sim 1/2$), however this level of precision is enough for our estimates. We see that the suppression for the annihilation rate at small $\alpha$ is much stronger than for transitions.

For higher $l$ levels the suppression at small $\alpha$ becomes even stronger, because the multipole number of the emitted graviton grows with $l$.
For instance, the annihilation rate for the $3d$ level scales as $\alpha^{20}$. Still, given that the power of $\alpha$ is already very high even for the $p$-level annihilation and the suppression in the annihilation rate is determined by how non-relativistic axions in the cloud are,
one may expect the annihilation rates for different levels to be comparable at equal values of the axion velocity, or, equivalently, at equal
values of $\alpha/l$. To illustrate that this is indeed the case, we present in Fig.~\ref{annihilation} the annihilation rates  for the first three superradiant levels, $l=1,2,3$, in units of $r_g$ for a cloud mass of $10^{-4}~M{\text{\tiny{black hole}}}$ as a function of $\alpha/l$. We see that they indeed agree within an order of magnitude.

\subsection{Axion non-linearities}
\label{nonlinear}
Let us turn now to another important source of nonlinear terms in (\ref{kineq})---self-interactions of the axion field itself.
As we discussed, the superradiant cloud is always close to be non-relativistic, so let us first discuss  the self-interaction effects in the non-relativistic approximation. In the non-relativistic limit, the axion field takes the form
\be
\label{smallv}
\phi={1\over\sqrt{2\mu_a}}\l e^{-i\mu_a t}\psi+e^{i\mu_a t}\psi^*\r\;,
\ee
where the characteristic scales for space and time variations of the function $\psi$ are much longer than $\mu_a^{-1}$. Then we plug the ansatz (\ref{smallv}) in the axion action and drop all rapidly oscillating terms. As the result we obtain the following effective action for $\psi$,
\be
\label{GrossPitact}
S_{nr}=\int d^4x\l i\psi^*\d_t\psi-{1\over 2\mu_a}\d_i\psi\d_i\psi^*-\mu_a\Phi\psi^*\psi+{1\over 16f_a^2}(\psi^*\psi)^2\r\;,
\ee
where $\Phi$ is the Newtonian gravitational potential of the black hole, and we kept only the leading non-linear term from the axion potential.
If we also drop the quartic self-interaction term in (\ref{GrossPitact}), then the equation following from (\ref{GrossPitact})  is the conventional Schroedinger equation in the external gravitational field. As before, the superradiant instability can be thought of as coming from an unconventional boundary condition at the origin. With the quartic term taken into account, the axion action (\ref{GrossPitact}) leads to the non-linear Gross--Pitaevskii
equation, well-known in condensed matter physics  to describe the dynamics of an interacting BEC (see, e.g., \cite{Pethick}
for an introduction). The sign of the interaction term in (\ref{GrossPitact}) corresponds to an attractive interaction between axions.
The black hole gravitational potential plays the role of the BEC trap.

Note, that as compared to gravity, axion self-interactions give rise to higher-order non-linearities (quartic, rather than cubic), however, they are suppressed by the scale $f_a$, which can be significantly lower than $M_{Pl}$. 

Let us describe the major consequences of these non-linearities.  We postpone the systematic discussion of how superradiant 
instability develops  till the next section~\ref{signatures}, but it is intuitively clear  that typically the axion field in the  cloud is dominated by a single harmonic in the expansion (\ref{field_harmonics})---the one corresponding to the fastest available superradiant level.
This is especially natural to expect at small and moderately large values of $\alpha$, when superradiant levels are very sparse.
So let's first consider how non-linearities affect a cloud composed of a single superradiant level.

\subsubsection{Bosenova}
As the number of axions in the cloud increases, the attractive force between axions becomes more and more important and
at some point the shape of the cloud changes significantly as compared to the one corresponding to the unperturbed hydrogen wave functions. We can estimate when this happens by equating the potential energy of axions in the cloud to the self-interaction energy,
\be
\label{varest}
{\alpha\over r}\sim {\psi^*\psi\over 8f_a^2}\;.
\ee
By integrating (\ref{varest}) over the volume we obtain that self-interaction effects become important in determining the shape of the cloud when
\[
N\gtrsim16\pi\alpha f_a^2 r_c^2\sim 16\pi{l^4\over \alpha^3}f_a^2r_g^2\;,
\]
where $N$ is the number of axions. Here we made use of (\ref{cloudsize}) for the size of the cloud, and set $\bar n\sim l$. It is more convenient
to write this bound as a condition on the mass $M_a$ of the axion cloud,
\be
\label{masscond}
{M_a\over M_{BH}}\gtrsim 2{l^4\over \alpha^2}{f_a^2\over M_{Pl}^2}\;.
\ee
We see that for typical values of the parameters we are interested in, $f_a\sim M_{GUT}$, non-linearities start playing an important role in determining the shape of the cloud quite early---when the cloud constitutes only $10^{-4}\div 10^{-3}$ of the black hole mass, and even earlier if the axion scale $f_a$ is significantly below the GUT scale.

As the cloud grows and its size is close to saturating (\ref{masscond}), the shape of the cloud is deformed and is no longer determined by the hydrogen wave functions.
However, a much more dramatic effect happens as the size of the cloud keeps growing. The effect was experimentally observed in trapped BEC's with attractive interactions and is known under the name ``Bosenova". Above some critical mass of the cloud the gradient energy of the axion field (``quantum pressure") cannot compete with the attractive force due to self-interactions and the cloud collapses. Indeed, it follows from (\ref{GrossPitact}) that the energy of the static cloud has the following parametric dependence on its size $r$,
\be
\label{novapot}
V(R)\simeq N{l(l+1)+1\over 2\mu_a r^2}-N{\alpha\over r}+{N^2\over 32\pi f_a^2 r^3}\;.
\ee
At small $N$ this energy has a minimum corresponding to a (meta)stable cloud, however, at large $N$ the last term in (\ref{novapot}) dominates over
the repulsion due to the quantum pressure term, and the cloud collapses.

\subsubsection{Shutdown of superradiance due to level mixing}
Yet another important consequence of non-linearities is that they may stop further development of superradiance. To see how, let us first consider the Press--Teukolsky ``black hole bomb"---a rotating black hole surrounded by a spherical mirror. As we discussed in the Introduction, this system provides the simplest example of a superradiant instability---a single photon introduced inside the mirror and satisfying the superradiance condition (\ref{Omegacond}) bounces between the mirror and the horizon and gets amplified in the ergo-region. Now imagine that the mirror has a defect and does not possess a perfect rotational symmetry around the axis of black hole rotation.
Then, when scattering off the mirror, some of the photons satisfying the superradiant condition (\ref{Omegacond}) change their quantum numbers. As a result they may exit the superradiant regime and be absorbed at the horizon. If the defect is substantial enough this may dump the superradiant instability.

Coming back to axion superradiance, as a consequence of self-interactions the axion cloud itself acts as a defect and may dump the further development 
of the instability when it becomes large enough. To analyze this at a more quantitative level, let us write the axion wave function in the form
\[
\psi=\psi_0+\delta\psi\;,
\]
where $\psi_0\propto e^{-i\l \omega_0t-m_0\varphi\r}$ is the field of the axions populating the most occupied level, and $\delta\psi$ is a perturbation.
We assume that the time scale for the growth of the cloud is much longer than the oscillation period for $\psi$ and consider the dynamics at scales short compared to the instability time---under these assumptions $\omega_0$ can be taken real. 
Then the linearized equation for the perturbation $\delta\psi$ has the following form, 
\be
\label{deltapsieq}
i\d_t\delta\psi=-{\d_i^2\delta\psi\over 2\mu_a}+\mu_a\Phi\delta\psi-{1\over 8 f_a^2}\l2\psi_0^*\psi_0\delta\psi+\psi_0^2\delta\psi^*\r\;.
\ee
The interesting feature of this equation is that it mixes $\psi$ and $\psi^*$; this property is typical for BEC perturbations and gives rise to the notion of Bogoliubov's quasiparticles. Namely, the solution to (\ref{deltapsieq}) mixes postive and negative frequency components,
\be
\label{deltapsisol}
\delta\psi=e^{-i\l \omega_0t-m_0\varphi\r}\l u(r,\theta)e^{-i\l \delta \omega t-\delta m\varphi\r}-v^*(r,\theta)e^{i\l\delta \omega t-\delta m\varphi\r} \r\;.
\ee
In the next section we will be interested in a situation, when the unperturbed cloud has parameters very close to the boundary of the superradiant region,
$\omega_0=m_0 w_+$ and the perturbation $\delta\psi$ corresponds to the fastest available superradiant level, which is $l=m=m_0+1$. This same reasoning applies for the levels with $m$ higher than $m_0+1$. The first term in (\ref{deltapsisol}) would give rise to such a perturbation, however, we see that as a result of the interaction with the background BEC, the perturbation has also 
an admixture of the non-superradiant $m=m_0-1$ modes. Of such modes the one with the fastest damping rate also has $l=m_0-1$, and as far as we can tell, there is no reason that would forbid an order one overlap of the function $v$ with the $l=m=m_0-1$ mode. 

\begin{figure}[t] 
 \begin{center}
 \includegraphics[width=6in,trim=70 70 50 100]{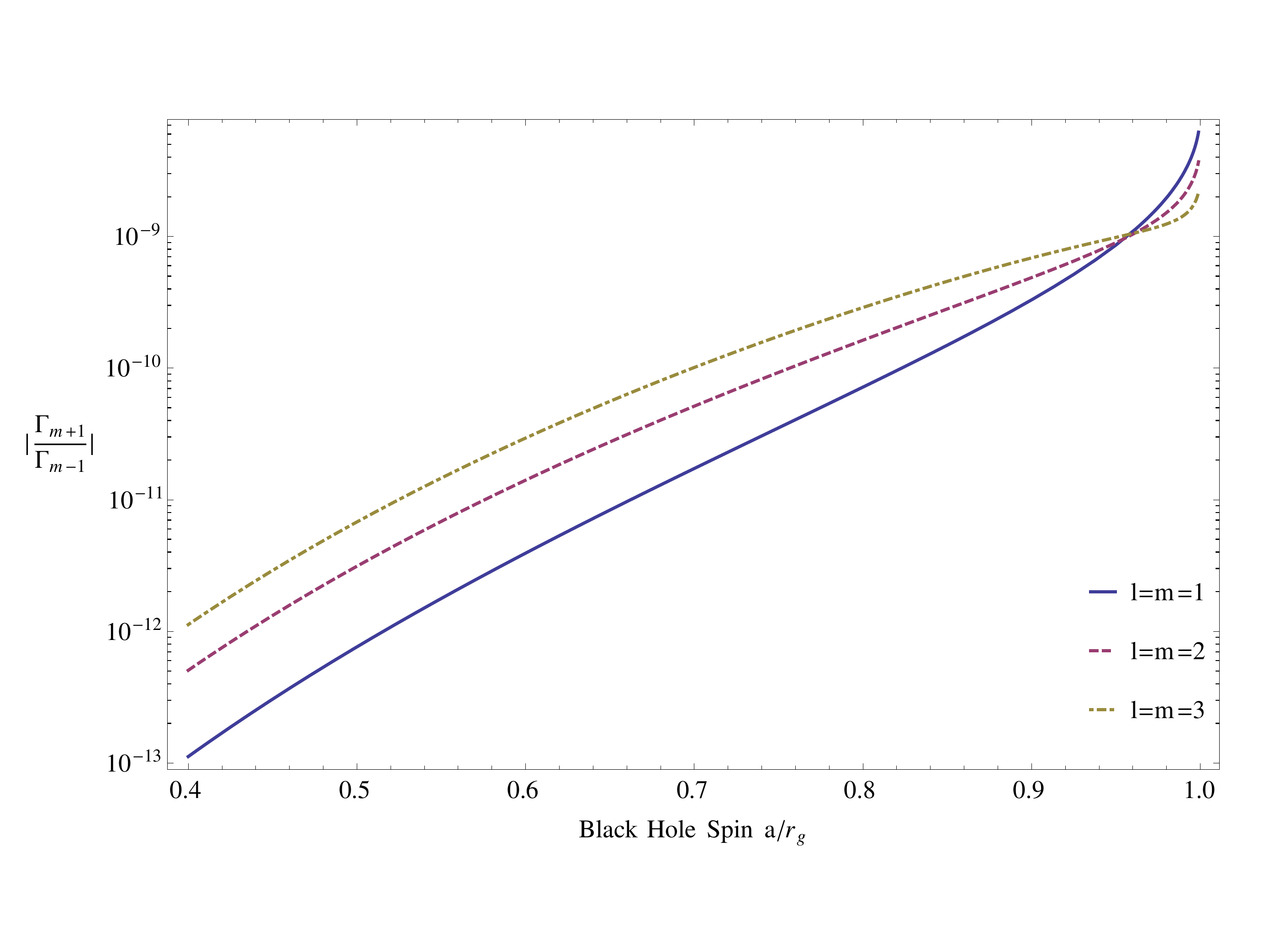}
 \caption{Ratios of the superradiance rate for the $(l+1)$ level to the absorption rate for  $(l-1)$ level if  the $l$ level has a vanishing imaginary part for $l=1, 2, 3$.}
 \label{Gammaratio}
 \end{center}
\end{figure}

To see whether the perturbation (\ref{deltapsisol}) is superradiant or dumped, let us proceed as we did before in the derivation of the time averaged energy flux through the black hole horizon, (\ref{horizonflux}). The axion field has now three different harmonics with frequencies $\omega_0$, $\omega_0\pm\delta \omega$. However, only the latter two contribute to the flux, because the first one saturates the superradiance condition (and all cross-terms vanish as a result of time-averaging).
As a result, the flux takes the following form
\be
\label{horizonfluxuv}
\langle P_\mu{\cal G}^\mu\rangle=\omega_1(\omega_1-(m_0+1)w_+)|u_h|^2+\omega_2(\omega_2-(m_0-1)w_+)|v_h|^2\;,
\ee
where $\omega_{1,2}=\mu_a+\omega_0\pm \delta \omega$, and $u_h$ and $v_h$ are the values of the functions $u$ and $v$ at the black hole horizon. For our choice of parameters, the $u$-term in (\ref{horizonfluxuv}) gives rise to the energy flux from the black hole, while the $v$-term gives rise to the flux into the black hole.
To deduce the direction of the net energy flux, let us recall that derivation of the WKB formula (\ref{GammaWKB}) implies, that the  functions $u_h,v_h$ at the horizon are related to their values  $u_c,v_c$ at the 
location of the cloud as,
\[
{|u_h|^2\over |v_h|^2}\sim\left| {\Gamma_1\over\Gamma_2}\right|{|u_c|^2\over |v_c|^2}\;,
\]
where $\Gamma_1$ and $\Gamma_2$ are superradiance and dumping rates for the two levels. 
 In turn, the ratio $u_c/v_c$ is determined by the relative strength of the non-holomorphic in $\delta\psi$ term in the perturbed Gross--Pitayevskii equation
 (\ref{deltapsieq}),
 \[
 {v_c\over u_c}\sim {\psi_0^2\over 8 f_a^2\mu_a\Phi}\sim {N\over 8\alpha f_a^2 r_c^2}\;.
 \]
 Combining these two relations together, we obtain that the perturbation $\delta\psi$ is superradiant if
the number of axions in the cloud is smaller than
\[
N\lesssim\left|{\Gamma_1\over\Gamma_2}\right|^{1/2}16\pi\alpha f_a^2 r_c^2\;.
\]
As in (\ref{masscond}) it is convenient to rewrite this condition as a bound on the fractional mass of the axion cloud,
\be
\label{strongmasscond}
{M_a\over M_{BH}}\lesssim  \left|{\Gamma_1\over\Gamma_2}\right|^{1/2}2{l^4\over \alpha^2}{f_a^2\over M_{Pl}^2}\;.
\ee
To estimate the ratio of rates in (\ref{strongmasscond}) we can use (\ref{Detweiller}). 
In Fig.~\ref{Gammaratio} we plot the result for the first few values of $m_0$ as a function of $a/r_g$. We see, that for a broad range of $a/r_g$ this ratio changes
between $\sim 10^{-12}$ and $\sim 10^{-9}$.  Note that for $a/r_g$ close to one the non-relativistic approximation is not accurate for the $l=m_0-1$ level.
Comparing with the numerical results of \cite{Dolan:2007mj} for $m_0=1,2$ suggests that the ratio $|\Gamma_1/\Gamma_2|$ is actually close to $10^{-9}$ at $a/r_g$ close to one, rather than to $10^{-8}$ as shown in Fig.~\ref{Gammaratio}.

By comparing (\ref{strongmasscond}) and (\ref{masscond}), we conclude that a large occupation number for one of the superradiant levels may indeed strongly inhibit the development of superradiance for other levels even in the regime when non-linearities are still irrelevant for the shape of the cloud. 

Importantly, this does not happen for the most occupied level itself---this corresponds to considering $\delta w=\delta m=0$ in (\ref{deltapsisol}).
The derivation of the horizon flux (\ref{horizonflux}) did not assume that the field is linear, so that as soon as the field is well approximated by
a single exponent $\psi_0\propto e^{-i\l \omega_0t-m_0\varphi\r}$ the superradiance will continue even in the non-linear regime. 

It is important for this argument that the ansatz $\psi_0\propto e^{-i\l \omega_0t-m_0\varphi\r}$ is a consistent solution of  the Gross--Pitaevskii equation at the non-linear level. This is no longer true if one considers the full scalar equation including all relativistic corrections---the higher harmonics get generated. However, we do not expect those to change the conclusion. As shown in section~\ref{gravemission}, relativistic processes are strongly suppressed for superradiant levels. Still, this point deserves further study.

Note, that the effects discussed so far---deformation of the shape of the cloud and shutdown of superradiance due to level mixing---in principle could be caused by gravitational backreaction of the cloud as well. We did not discuss them in  section~\ref{gravemission}, because the shape deformation is always small as soon as $M_a\ll M_{BH}$, and the level mixing is absent in the non-relativistic limit when the density of axions $\rho_a\approx\mu_a\psi^*\psi$ does not depend on the azimuthal angle  $\varphi$, as the cloud is dominated by a single level. Consequently, these gravitational effects are likely to be always subdominant with respect to those caused by axion self-interactions.

Conversely, axion self-interactions may cause effects similar to those discussed in section~\ref{gravemission}, annihilations and elastic scatterings of axions. The leading annihilation process is annihilation of three axions from the cloud into one axion in the continuum. Given that the outgoing axion has energy of order $3\mu_a$, in the leading approximation it can be considered as massless, and the calculation of the emission rate can be done similar to the graviton case. The analogue of (\ref{weinberg_power}) for three axion annihilation in the massless approximation reads
\be
\label{scalar_power}
{dP\over\sin\theta d\theta d\varphi} =2\omega^2|j(\omega,k)|^2\;,
\ee
where
\[
j(\omega,k)\equiv {1\over 4\pi}{\mu_a^2\over 6f_a^2}\int d^3{\bf x}\phi^3(\omega,{\bf x})e^{-ik{\bf x}}\;.
\]
Then a straightforward calculation gives the following result for the annihilation of three axions from the $l=m=1$ level  into a single unbound  axion,
\be
\label{3to1}
{dP\over\sin\theta d\theta d\varphi}(3\times 1p\to continuum)\approx N^3 {2^{10} \alpha^{23}\over 3^{10}f_a^4\pi^3r_g^6(4+\alpha^2)^{10}}\sin^6{\theta}\;.
\ee
To compare the efficiency of this process to the one graviton annihilation (\ref{11annihilation}),  let us integrate over the angles in both cases
to calculate the  total emitted power and take the ratio. We  get
\be
\label{axgravratio}
{P(3\times 1p\to continuum)\over P(2\times 1p\to graviton)}\approx 10^{-2}\alpha^4{M_a\over M_{BH}}{M_{Pl}^4\over f_a^4}\;,
\ee
so that self-interactions dominate when the size of the cloud is not too small, however, as the cloud decreases, the two axion annihilation into gravitons takes over. For instance, it will typically be more important when the cloud size approaches the value in (\ref{strongmasscond}).

Just as in the case of graviton emission one may look for the processes that conserve axion number and as a result may be less suppressed by powers of the small  axion velocity in the cloud. An obvious candidate process is an elastic two-to-two scattering of axions. However, it appears likely that such processes are not important. The reason is that for the two-to-two scattering to be unsuppressed, three of the participating axions should correspond to highly occupied levels. As we will discuss in section~\ref{signatures}, typically at any given moment of time only few of the levels are significantly occupied, and usually there is one which dominates the cloud. Then the most likely candidate for the scattering process is a scattering when two axions from the most populated level scatter, and one goes down to another highly populated level, while the remaining axion flies out in the continuum. For this scattering to be compatible with the energy conservation one needs
\[
{2\over \bar{n}_1^2}< {1\over \bar{n}_2^2}\;,
\]
where $n_1$ and $n_2$ are the principal quantum numbers of the high- and low-lying levels, respectively. The discussion of the superradiance development in the next section implies that the situation where two levels satisfying this condition simultaneously have large occupation numbers is hardly possible.

Note that throughout most of the discussion of axion self-interactions and gravitational wave emission we treated axions in the cloud as free particles with hydrogen wave-functions, while strictly speaking the elementary excitations of the axion BEC are Bogoliubov quasiparticles (\ref{deltapsisol}). The free particle approximation is accurate when the mass of the cloud is small, so that  the effects of self-interactions are weak, but may be misleading when the mass of the cloud is close to saturating the bound (\ref{masscond}).
Unfortunately, it's hard to improve on this approximation without going into numerical simulations of the cloud, which are beyond the scope of our paper. We proceed under assumption that the free particle approximation is a reliable guide for an order of magnitude estimates at the masses close to saturating the bound  (\ref{masscond}) as well. The level mixing phenomenon described above provides an important example of a situation, when the free particle approximation is not adequate even at very small masses of the cloud. There is a clear physical reason why this happens---some of the levels have damping rates orders of magnitude faster than the population rates for the relevant superradiant levels, so even a tiny level mixing qualitatively changes the dynamics. Fortunately, this effect is straightforward to take into account perturbatively, as we did.

To summarize, this discussion implies that self-interaction effects cause a strong influence on the phenomenology of superradiance. 
The proper taking into account of these effects is one of the major challenges for obtaining an accurate quantitative description of superradiance development. The estimates presented here are far from being a complete accurate treatment and it appears likely that numerical simulations are required to really solve the problem. It is worth mentioning that level mixing can also be caused by the accretion disk of the black hole or a massive object orbiting the black hole but we have already shown in \cite{Arvanitaki:2009fg} that these can be safely ignored. We proceed now with estimating the possible observational signals of superradiance.
   
\section{Observational signatures}
\label{signatures}

Now that we are well equipped with the details of how superradiance works, let us put them together and develop an intuition about the way superradiance develops in realistic environments and about the observational signatures we may expect from this process under various circumstances. The full treatment of the set of kinetic equations describing superradiance (\ref{kineq}), (\ref{massevol}), (\ref{spinevol}) appears to be quite challenging, given that a large number of competing processes with drastically varying time scales is involved. Our strategy will be to start with a highly  idealized situation including  a minimal number of dynamical ingredients and then keep adding more processes to get closer to a realistic description. We already presented the list of possible observational signatures of superradiance in the {\it Introduction}. Clearly, following the above strategy the very first signature to discuss is the absence of rapidly rotating black holes of size matching  the axion Compton wavelength---the black hole spin-down is the most direct consequence of superradiance.
\subsection{Black hole Regge trajectories}
To get a rough idea of the bound on the axion mass that could come from black hole spin measurements,  we present in Fig.~\ref{aeq1plot} regions in the black hole mass vs axion mass plane where the superradiance timescale for a maximally rotating black hole is shorter than  the age of the Universe and the Eddington accretion time. We used the superradiance rates of section~\ref{spectroscopy} to produce this plot. For superradiance to have a non-negligible effect on the black hole spin the process should last for many $e$-foldings for the produced axions to carry away a noticeable fraction of the black hole spin. To estimate the required number of $e$-foldings, note that, approximately,
\be
\label{axionnumber}
{M_{BH}\over \mu_a}={M_{BH}r_g\over \alpha}\sim 10^{76} \l {M_{BH}\over M_{\odot}}\r^2
\ee
of axions need to be produced for their total mass (spin) to be of order the black hole mass (spin). This requires  $\sim10^2$ $e$-foldings of superradiance; we took this factor into account in Fig.~\ref{aeq1plot}, by presenting the region where the age of the Universe (or Eddington accretion) time is longer than hundred superradiance times. We see from the plot that, as the black hole mass grows, the size of the interval in axion mass which the black hole can in principle probe shrinks, because the superradiance time gets longer. 

This plot is useful as a zeroth order estimate, but cannot be used to deduce limits on axions from data on existing black holes, since the black holes being observed do not all have spins close to the maximum. Instead, we need to know the regions in the black hole ``Regge" plot, the spin vs mass plane for black holes, where we do not expect to find black holes if an axion with a certain mass exists, as shown in Fig. \ref{fig_summary}. The intricate structure of these plots is a manifestation of level quantization in the superradiant gravitational atom, and their underlying physics is explained in what follows.


To start with, let us ignore accretion and consider a black hole that starts off as maximally rotating. This approximation should be physically relevant for stellar mass black holes produced as a result of a fast catastrophic event, such as the supernovae explosion. Deviations from spherical and axial symmetry are believed to be crucial for supernovae explosions, so there should be a lot of angular momentum available when the black hole forms, and one may expect high initial values for the black hole spin---as soon as the supernovae core gets rid of all the angular momentum above the extremal value it collapses and forms a rapidly rotating black hole. This expectation seems to be supported by observations---for instance, the high value of the spin-to-mass ratio $\frac{a}{r_g}\approx 0.92^{+0.05}_{-0.07}$ deduced \cite{GouMcClintock} for the black hole primary in the extragalactic X-ray binary LMC X-1 is hard to reconcile with the young age ($\sim 5\cdot 10^6$~yr) of the system, if the spin were not natal.
 
For concreteness, we assume that the parameter $\alpha$ for the black hole is small, $\alpha\lesssim 1/2$, so initially the fastest superradiant level is the $2p$ level with $l=m=1$. Then, initially one can ignore all the levels apart from the $2p$ superradiant level. It is straightforward to generalize all the discussion below to smaller initial values of the black hole spin and larger values of $\alpha$. The black hole will start to lose its spin rapidly by  populating this level. The time scale for this process can be really fast---from section~\ref{spectroscopy} we know that the superradiance rate can be as fast as $10^7 r_g^{-1}$,
which corresponds to $10^2$ seconds for a two solar mass black hole. However, there is a critical value of the black hole spin $a_1(\alpha)/ r_g$,  at which the superradiant condition (\ref{Omegacond}) ceases to hold for the $l=m=1$ level, so the width of this level becomes zero and  the spin-down process  terminates. 

Let us for a moment consider the case where the axion self-interactions are absent, as if we were dealing with a free massive field rather than a (pseudo)Nambu-Goldstone boson. Then at this point superradiance would continue by populating the second superradiant level $l=m=2$ ($3d$) at a much longer time scale. Note, however, that for many $e$-foldings of superradiance the black hole spin would remain approximately constant and equal to $a_1(\alpha)/ r_g$. Indeed, if the spin would significantly drop below this value the frequency of the $2p$-level would acquire a {\it negative} imaginary part, so that the black hole would start
absorbing axions from the $2p$-level and spinning up back with a rate much faster than the population rate for the $3d$ level. Instead, the spin stays practically constant close to 
$a_1(\alpha)/ r_g$
 as the black hole populates the $3d$-level while being slowly fed by axions from the $2p$-level.  When the occupation number for the latter level $N_{2p}$
 drops below $\frac{\Gamma_{3d}}{ \left| \Gamma_{2p}\right|} N_{3d}$ the spin-up rate due to $2p$-level cannot compete with the spin-down rate due to $2d$-level and the black hole spin further drops down till the value $a_2(\alpha)/ r_g$, where the superradiance rate for the $3d$-level turns zero and the story repeats this time involving the $l=m=3$ ($4f$) level.

Of course, from section~\ref{dynamics} we know that this story cannot be an accurate description of what actually happens---non-linearities related to the axion self-interactions and due to gravity cannot be neglected in a realistic description of superradiance. However, the above simplified example correctly captures the major important feature---during the spin-down black hole spin tends to evolve rapidly till it reaches the line $a_i(\alpha)/ r_g$
in the Regge plane where one of the superradiant levels changes the sign (``Regge trajectory"), where it can get stuck for a quite long period of time.
In fact, as we will see momentarily, non-linear effects make this behavior even more pronounced.

Indeed, as we discussed in section~\ref{nonlinear}, even relatively small amount of axions populating one superradiant level may shut down the instability
for the next level. For instance, in the above example, when the black hole reaches the first Regge trajectory $a_1(\alpha)/ r_g$ the $3d$ level does not start being populated until a large enough number of axions  dissipate from the $2p$ level, so that its mass drops below the bound in (\ref{strongmasscond}).

\begin{figure}[t] 
 \begin{center}
 \includegraphics[width=7in,trim=0 70 0 100]{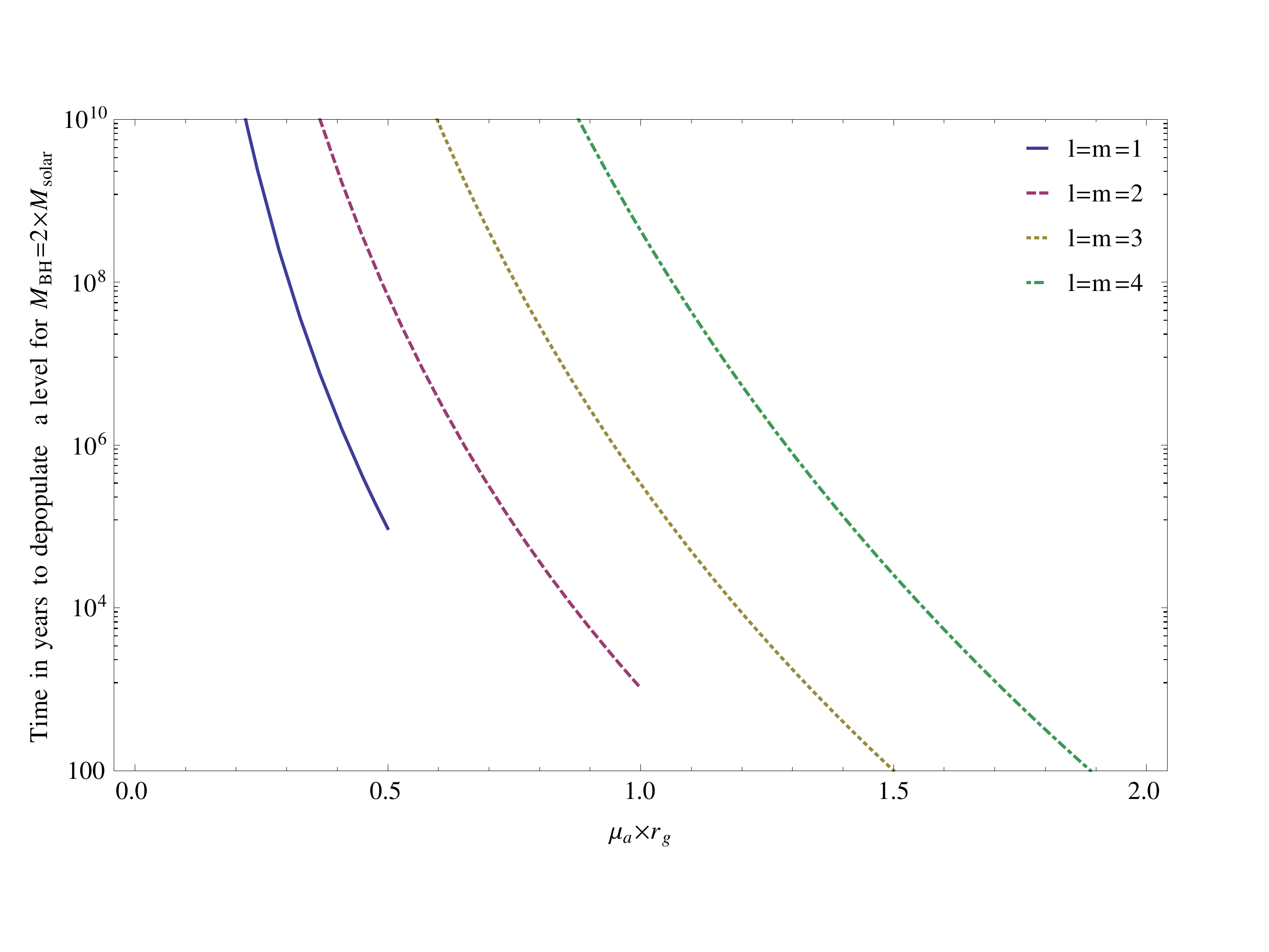}
 \caption{The time required for the axion cloud around a $2M_\odot$ black hole to dissipate such that the next superradiant level 
 can start being populated for clouds with $l=1, 2, 3, 4$.}
 \label{fig:cleanup}
 \end{center}
\end{figure}

We discussed two processes that reduce the number of axions in the superradiant cloud, annihilations into gravitons and annihilations into unbound axions due to self-interactions. The latter process is more efficient at large occupation numbers. However, the annihilation rate drops down as the number of axions decreases and, when the cloud mass approaches the bound (\ref{strongmasscond}), the graviton annihilation, which involves only two axions, wins, as seen from (\ref{axgravratio}).
The total duration of the annihilation period before superradiance restarts, is dominated by the latest stages of the process. We can estimate the duration of this period by using the annihilation rates calculated in
section~\ref{gravemission}. The occupation number of axions dissipating from the cloud through annihilations into gravitons satisfies the following equation
\be
\label{anniheq}
{d N\over dt}(2\times axion\to graviton)\equiv -\Gamma_{ann}N^2\;,
\ee
where the coefficient $\Gamma_{ann}$ can be deduced by integrating the annihilation rates, such as (\ref{11annihilation}), over the angles.
By solving (\ref{anniheq}) we obtain that the occupation number evolves in time as
\be
\label{Nevol}
N(t)={N(0)\over 1+\Gamma_{ann} N(0)t}\approx {1\over \Gamma_{ann} t}\;,
\ee
where at the last step we took the late time asymptotics. By using (\ref{strongmasscond}) we find that the annihilation time needed to clean the system before  superradiance can continue to populate the next level is
\be
\label{cleanup}
t\sim{\alpha^2 \mu_a\over2l^4M_{BH}\Gamma_{ann}}{M_{Pl}^2\over f_a^2}\left|\Gamma_2\over \Gamma_1\right|^{1/2}\equiv \tau(\alpha) \l{M_{BH}\over 2M_\odot}\r\l{ M_{Pl}^2\over f_a^2}/10^4\r\l\left|\Gamma_2\over \Gamma_1\right|/10^{-10}\r^{1/2}\;,
\ee
where the last step is merely the definition of the normalized annihilation  time $\tau(\alpha)$.
In Fig.~\ref{fig:cleanup} we present $\tau(\alpha)$ for the first three levels. We see that for stellar mass black holes, depending on the parameters, there is enough time for annihilations to complete on one or two Regge trajectories.

The above discussion gives rise to the following picture of the black hole evolution under the influence of superradiance. An initially fastly rotating black 
rapidly loses its spin and approaches the closest Regge trajectory $a_i(\alpha)/r_g$. Then, for a long time the black hole stays at this trajectory while axions in the cloud keep annihilating. When the cloud mass drops below the critical value (\ref{strongmasscond}) the superradiance becomes operative
again and the black hole rapidly travels to the next Regge trajectory. Consequently, in the region of the Regge plane affected by superradiance masses and spins of observed black holes should follow quantized trajectories.

During transitions between Regge trajectories  another non-linear process discussed in section~\ref{nonlinear}---bosenova---becomes important. As the cloud mass during the transition grows above (\ref{masscond}) the cloud becomes unstable as a result of the attractive axion self-interactions and collapses. The detailed analysis of this process requires numerical simulations, which are beyond our goals in this paper. However, the most likely outcome seems to be that order one fraction of the cloud gets absorbed by the black hole and order one becomes relativistic and escapes at the time scale set by the size of the black hole. Condition (\ref{Omegacond}) implies that to complete the transition to the next Regge trajectory a black hole needs to release up to a few percent of its mass into axions. Consequently, each transition proceeds through a sequence of tens to hundreds of Bosenova explosions for $f_a\sim M_{GUT}$. As discussed later, these explosions may give rise to the observable gravitational wave signal for supermassive black holes and perhaps to the gamma ray signal for the QCD axion.

It is straightforward now to add accretion into this picture, at least at the qualitative level. Under the influence of accretion, the black hole mass and spin will still stay on the Regge trajectory, since the superradiance rate for the corresponding level is much faster than the accretion rate away from the trajectory. Indeed, if accretion brings the black hole above the Regge trajectory, the level acquires positive imaginary part and spins the black hole down back onto the trajectory. Conversely, if as a result of accretion the black hole deviates below the Regge trajectory, the imaginary part becomes negative and the black hole starts absorbing axions from the cloud to return on the trajectory.

Note that accretion may affect the black hole transition rate between different trajectories if its rate is faster or comparable to the annihilation rate. If it consistently pushes black hole above the  trajectory, new axions will be coming to the cloud compensating the effect of annihilations. Conversely,
by pushing the black hole below the trajectory, accretion may accelerate the dissipation of the cloud.

Of course the above discussion only applies if accretion is slow compared to superradiance in the vicinity of the Regge trajectory. We illustrated all of the above in Fig.~\ref{fig_summary}. Here, lines of different colors correspond to different levels. Parts of these lines where the spin increases with the mass
are boundaries of the superradiant region, where $\alpha=mw_+(a/r_g)$ for the corresponding values of $m$. These are the Regge trajectories $a_i(\alpha)/r_g$ discussed above. For each trajectory  the superradiance time grows at small $\alpha$, and at some point becomes longer than the age of the Universe. Then, instead of showing the line where the width of the level is zero, we show the curve where the superradiance time is equal to the age of the Universe. These are the parts of the lines in Fig.~\ref{fig_summary}, where the spin is a decreasing function of the mass. 
 Below the solid line the superradiance time is longer than the age of the Universe for all unstable levels.

To finish the discussion of the Regge trajectories, note that we started with considering an example of the black hole spin-down, which can be relevant for stellar mass black holes, but by now it is clear that also the evolution of the galactic black hole will follow the same rule---as the black hole enters in the region of the Regge plane affected by superradiance it starts moving there along the Regge trajectories, occasionally experiencing rapid transitions between different Regge trajectories.

\subsection{Gravitational waves}
\label{wavesignature}
An even more direct possibility to detect the presence of an axion cloud around black holes is to observe the associated gravitational wave signal. 
%
%
As we discussed in section~\ref{gravemission} there are two principal processes giving rise to gravitational waves from the cloud---axion transitions 
between levels and axion annihilations. The bosenova collapse may also give rise to a burst of gravitational waves.

Let us start with the transition signal.  For  transitions  to be efficient one needs large occupation numbers for two different levels to be present simultaneously to get a Bose enhancement of the signal. Combined with non-linear effects discussed in section~\ref{nonlinear} this practically singles out the type of transitions having chances to be observed and the corresponding stages of the black hole evolution\footnote{In particular, transitions between superradiant and non-superradiant levels, chosen as an illustrative example in \cite{Arvanitaki:2009fg}, are actually never important because the transition rate is always suppressed compared to the axion absorbtion rate for non-superradiant levels and they never have a chance to acquire a large occupation number.}.

Indeed, in section~\ref{nonlinear} we found that even a relatively small amount of axions on the most populated level shuts down the superradiance for levels with different magnetic angular numbers as a result of level mixing. Consequently, the only chance for two levels to grow together, and as a result to acquire large occupation numbers 
 simultaneously, is when the levels have equal angular numbers $m$. 
This case corresponds to setting $\delta m=0$ in (\ref{deltapsisol}); the mixing for such levels does not change their magnetic number and does not shut down the superradiance. 

Furthermore, for two levels with
equal magnetic numbers $m$ but different orbital momenta $l$ the superradiance rate for the more energetic level (the one with a higher $l$) is very much slower, so that by the moment the black hole reaches the corresponding Regge trajectory by populating the lower level, the occupation number for the higher level is tiny, and there is no significant transition signal.

All this lead us to consider the transition between two levels with different principal quantum numbers $n$, but equal $l$ and $m$ as the most promising source of an observable gravitational wave signal. It is natural to consider the case, when the level with the larger principal number $n$ has 
the faster superradiance rate.
 As mentioned in section~\ref{lowalpha}, the lowest $l$ when such a situation takes place is $l=4$, 
 so let's pick this level as the simplest representative example.
 The corresponding transition rate is given by (\ref{grate}). The amplitude of  the gravitational wave signal at the detector is related to the total power emitted as
 \be
 \label{hformula}
 h=\l {4G P\over r^2 \omega^2}\r^{1/2}\;,
 \ee
 where $\omega$ is the frequency of emitted gravitons and $r$ is the distance to the source.
 By making use of the rate (\ref{gpower}) and plugging in the transition frequency (\ref{deltaw}) we obtain
 \be
 \label{htrans}
 h\sim 10^{-22}\alpha^2(\epsilon_1\epsilon_0)^{1/2}\l{10 \mbox{ Mpc}\over r}\r\l{M_{BH}\over 2M_\odot }\r\;,
 \ee
 where $\epsilon_{1,0}$ are total masses of axions populating the $n=1$ and $n=0$ levels, in units of the black hole mass,
 \[
 \epsilon_{1,0}={M_{a1,0}\over M_{BH}}
 \]
 and the frequency of the signal $\nu$ is determined by the axion and black hole masses as
 \be
 \label{tranfreq}
 \nu\approx100\,\mbox{Hz }\alpha^3\l{2M_\odot\over M_{BH}}\r\;.
  \ee

For annihilations one needs fewer conditions to be satisfied to get a significant signal---the occupation number for only one of the levels has 
to be large. As we discussed in section~\ref{gravemission} the annihilation rates for different levels are rather similar at same values of
$\alpha/l$, so let us consider the $l=1$ level as a representative example.
The annihilation rate (\ref{11annihilation}) gives rise to a gravitational wave signal of strength equal to
\be
\label{hann}
h\sim 10^{-22}\alpha^7 \epsilon \l{10 \mbox{ Mpc}\over r}\r\l{M_{BH}\over 2M_\odot }\r\;,
\ee
 where, as before, $\epsilon$ is the fraction of the black hole mass accumulated in the cloud. The frequency $\nu$ for this signal is given by 
 \be
 \label{annfreq}
 \nu\approx 30\,\mbox{kHz }\alpha\l{2M_\odot\over M_{BH}}\r\;.
 \ee

The numbers in (\ref{htrans}) and
(\ref{hann}) definitely appear interesting both at high frequencies probed by Advanced LIGO and corresponding to stellar mass black holes
and when scaled down to low LISA frequencies, corresponding to supermassive galactic black holes. However, to judge 
the chances to observe these signals we need to estimate  the fractional mass  of the cloud, $\epsilon$, in  (\ref{htrans}) and
(\ref{hann}), that will determine the possible observational reach in terms of the distance to the source. 
 
 To make these estimates note that,
 as we discussed, a large fraction of its time the black hole spends on the Regge trajectories with a relatively small axion cloud around,
 down to $\epsilon\sim10^{-12\div 10}$, waiting for the cloud to be dissipated so that the next level can start being populated triggering a relatively fast transition to the next Regge trajectory. Transitions between different trajectories and relatively short time intervals afterwards, when $\epsilon$ can be significantly larger, provide the most promising periods for an observable gravitational wave signal. Every black hole may experience several such transitions---one directly after the black hole formation, another one or two after periods of axion annihilations and possibly more triggered by accretion or merger events.
 
\begin{figure}[t!] 
 \begin{center}
 \includegraphics[width=6in,trim=70 20 50 20]{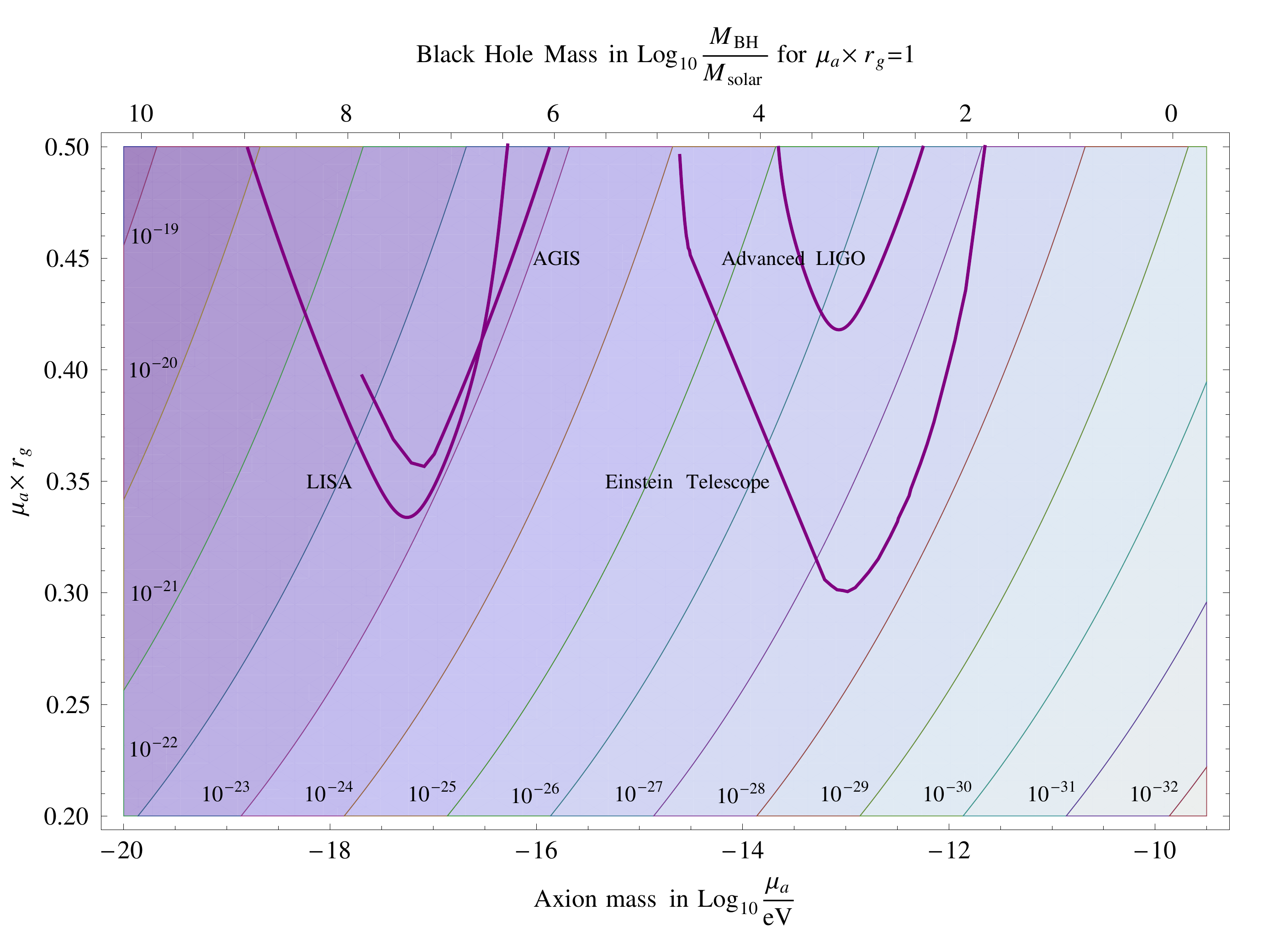}
 \caption{The contour plot of constant gravitational wave signal from axion annihilations in the $2p$ level for a black hole located at 20~Mpc away from the Earth. The projected sensitivity curves  assume $10^6$ seconds of a coherent integration time for LISA \cite{LISA2}, AGIS, Advanced LIGO and Einstein Telescope.}
 \label{annsignal}
 \end{center}
\end{figure}

From (\ref{Omegacond}) we can estimate the total spin, and, as a consequence, the mass extracted from the black hole during such transitions. The latter typically turns out to be  around a percent of the black hole mass. However, the main limiting factor for the size of the cloud is the Bosenova instability that prevents the cloud mass to grow above $10^{-4}\div 10^{-3}$ of the black hole mass.  This is enough to estimate the strength of the annihilation signal. In Fig.~\ref{annsignal} we present the contour plot for the strength of the signal in the black axion mass vs $\alpha$ plane from axion annihilations in the $l=1$ level, assuming the size of the cloud $\epsilon=10^{-4}$. We pick 20~Mpc as the distance to the source (which is the distance to the Virgo supercluster, containing about 2000 galaxies), and choose $10^6$ seconds as a coherent integration time for the signal. Note that existing spin measurements (Fig.~\ref{fig_summary}) suggest a lower bound of $\mu_a\gtrsim 3\cdot 10^{-11} $~eV for  axions matching the size of stellar mass black holes (which would correspond to $\alpha\gtrsim 0.9$ for $2M_\odot$ black hole), pushing the annihilation signal for Advanced LIGO into a range of somewhat uncomfortably large frequencies.

Predicting the characteristic  features of the transition signal is more involved and requires a detailed quantitative analysis of the dynamics, but some simple estimates  can still be done. Let us focus on the simplest case  of the $6g\to 5g$ transition. To describe the evolution of the cloud during the transition to the $l=4$ Regge trajectory, let us truncate the system  (\ref{kineq}) by keeping only $6g$ and $5g$ levels and ignore accretion. Then we obtain the following pair of equations for the relative sizes of the $6g$ and $5g$ components of the cloud,
 \begin{gather}
 \label{6g5g}
 {d\epsilon_0\over d t}=\Gamma_{440}\epsilon_0-\Gamma_t\epsilon_1\epsilon_0\\
 {d\epsilon_1\over d t}=\Gamma_{441}\epsilon_1+\Gamma_t\epsilon_1\epsilon_0\nonumber\;,
 \end{gather}
where $\Gamma_{440}$, $\Gamma_{441}$ are the superradiance rates, and the transition coefficient $\Gamma_t$ is determined from (\ref{grate}) to be equal to
\[
\Gamma_t\approx3\cdot 10^{-7}{\alpha^8\over r_g}\;.
\] 
We neglected the annihilation processes which are slow compared to superradiance and transitions. As follows from Fig.~\ref{rates} the superradiance rates $\Gamma_{440}$, $\Gamma_{441}$ are of order $10^{-10}r_g^{-1}$ for $\alpha\sim 1$. Using the small $\alpha$ approximation (\ref{Detweiller}) as a guide, their ratio can be estimated as 
$\Gamma_{440}/\Gamma_{441}\sim 0.9$.

Let us focus on the case of $\alpha\sim1$. Then the dynamics following from equations (\ref{6g5g}) is quite simple. Both levels start being populated but the lower one has a smaller superradiance rate, and as a consequence is less occupied. By the time the occupation number of the $6g$ level reaches its maximum, $\epsilon_1\sim 10^{-4\div3}$,
the occupation of the lower $5g$ level is given by 
\[
\epsilon_0\approx\epsilon_1 e^{-0.1 N_e}\;,
\]
where $N_e$ the number of e-foldings of superradiance required to populate the $6g$ level. The number of e-foldings depends on the initial number of axions. As follows from (\ref{axionnumber}) it varies from 
$N_e\sim 165$ if initially the $5g$ level is not occupied, down to  $N_e\sim 100$ if we estimate the initial occupation number to be determined by the dark matter density. In fact, 
the initial axion number can be significantly larger, if we consider a transition after a recent Bosenova event. 

\begin{figure}[t!] 
 \begin{center}
 \includegraphics[width=6in,trim=70 20 50 20]{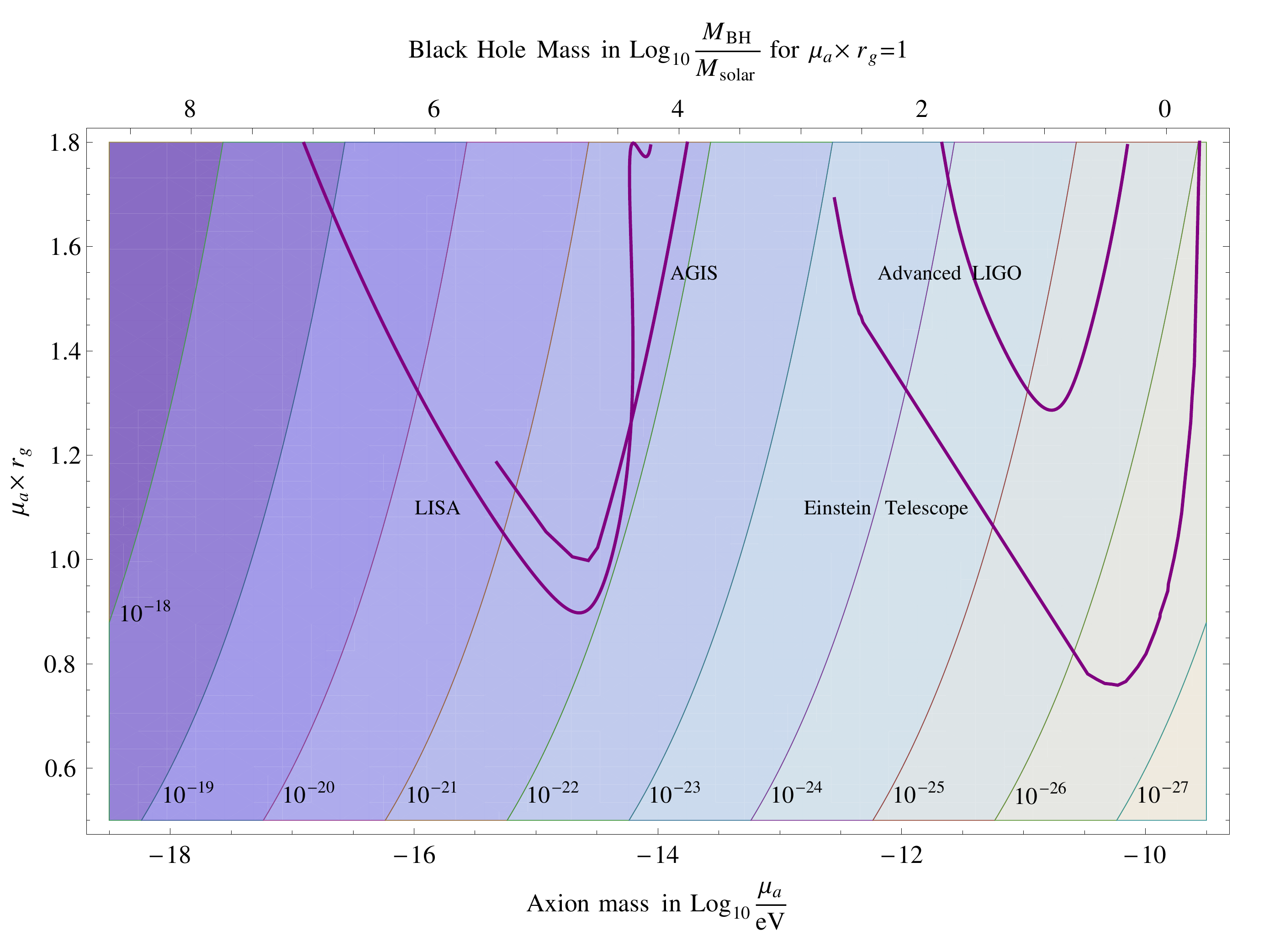}
 \caption{The contour plot of constant gravitational wave signal from axion transitions between the $6g$ and the $5g$ levels for a black hole located at 20~Mpc away from the Earth. The projected sensitivity curves  assume $10^4$ seconds of a coherent integration time.}
 \label{transignal}
 \end{center}
\end{figure}

Even though $\epsilon_0$ 
is exponentially sensitive to $N_e$ this uncertainty does not pose a big problem for estimating the gravitational wave signal at its maximum. Indeed, even if we set $\epsilon_0=\epsilon_1$, the transition terms in (\ref{6g5g}) are still too small to compete with the superradiance upto  $\epsilon_1\sim 10^{-4}$, so that this uncertainty does not affect the dynamics. Also, to estimate the maximal possible signal let us concentrate on the very last episode of spinning down which is  terminated because the black hole reaches the $l=4$ Regge trajectory (and not by the Bosenova event, as happens for earlier episodes of spindown). At the end of this episode superradiance shuts down and only the transition terms in (\ref{6g5g}) are left.
At this point  the $\epsilon_0/\epsilon_1$ ratio is small, but it starts growing as a result of transitions. The transition signal reaches its maximum when 
$\epsilon_0\sim\epsilon_1\sim 10^{-4}$ and then decreases because the occupation fraction $\epsilon_1$ for the $6g$-level drops down. The duration of the signal at
peak intensity is determined by the transition rate and is of order
\[
t(6g\to 5g)\sim 3\cdot 10^6{r_g\over \alpha^8\epsilon_0}\;,
\]
which is of order a day for a stellar mass black hole with $\alpha\sim 1$ and $\epsilon_0\sim 10^{-4}$.

In Fig.~\ref{transignal} we present the contour plot for the gravitational wave amplitude as determined by (\ref{htrans}) for different axion masses and values of $\alpha$ (equivalently, for different black hole masses), assuming $\epsilon_1=\epsilon_0=10^{-4}$ and taking 20~Mpc as a distance to the source.  We presented also the sensitivity curves of various planned gravitational wave detectors assuming the coherent integration time $10^4$ seconds. 

We see that future experiments will be sensitive to the transition signal from the superradiant cloud over a large range of axion and black hole masses.
What is especially exciting is that the Advanced LIGO detector which is scheduled to start operating around 2014 will be probing the heavy mass regime for axions, in particular the QCD axion. We will discuss  the Advanced LIGO reach for the QCD axion in more details in section~\ref{QCDaxion}.

Finally, as we already said, as a consequence of Bosenova, every transition between Regge trajectories goes through a series 
of 10-100 spin-down episodes
 interrupted by silent intervals needed to  build up the cloud. Depending on the distance to the source one may see also gravitational wave signal from the earlier episodes, although to study this possibility requires a more detailed analysis of the dynamics (in particular, accurate prediction of the e-folding number $N_e$). If these signals can be observed, then by 
 measuring the frequency and the amplitude of the signal as well as the duration of the active and silent intervals one may hope  to extract not only the mass of the axion, but to estimate the scale $f_a$ as well. 
 
 Also the Bosenova event by itself gives rise to a gravitational wave burst. Assuming that the collapse of the cloud happens on a time scale of order $r_g$
 the power emitted in gravitational waves during the Bosenova event can be estimated as
 \[
 P_{BN}\sim G_N \epsilon^2 M_{BH}^2r_g^{-2}\;,
 \]
which translates into the gravitational wave amplitude at the Earth of order
\[
h\sim\epsilon {r_g\over r}\sim 10^{-17}\l {\epsilon\over 10^{-4}}\r\l{M_{BH}\over 10^8 M_\odot}\r\l{100\mbox{ Mpc}\over r}\r\;,
\]
with the frequency being of order $r_g^{-1}$. This signal may be observable for supermassive black holes.

To conclude, let us emphasize that in estimating the signal strength we were using the free particle approximation for axions all the way up to $\epsilon\sim 10^{-4}$, when the bound  (\ref{deltapsisol}) gets saturated and this approximation may be not adequate, as we already discussed. This may be especially important for the transition signal where one needs to follow several levels simultaneously. Consequently, our encouraging estimates here should be considered as a strong motivation for a further careful numerical analysis of the system, rather than as an accurate prediction for the signal.

%
%
\subsection{Direct observation of the cloud}
\label{EMRI}
Another potential observational consequence of superradiance  is the presence of the cloud itself, which could be directly detected by precision mapping of the near horizon black hole metric. Such a mapping will be made possible by future low frequency gravitational wave detectors, such as LISA or AGIS, during the last stages of the inspiral of a compact object (black hole/neutron star/white dwarf) into a supermassive black hole. With LISA sensitivities, it will be possible to observe hundreds of such events per year for different galaxies, and to trace up to $\sim 10^5$ orbits in an individual  event. This will allow mass and spin determination with $10^{-5}$ accuracy, and about 6-7 higher multipole moments of the metric  can be measured with better than a few percent precision.

In principle, it is straightforward to calculate the modification to the waveform of the inspiral signal due to the presence of the axion cloud. In the regime when non-linearities can be neglected (and this is the only regime, where the cloud can stay  for a cosmologically long time) the shape of the cloud is determined by the well-known wave functions of the hydrogen atom. It is likely that the best chances to observe the presence of the cloud are for black holes with moderately small values of $\alpha/l$.
Indeed, at smaller values of $\alpha$ non-linear effects and processes leading to dissipation of axions from the cloud get suppressed allowing for longer lifetimes and a larger cloud mass. On the other hand, the size of the cloud grows at small $\alpha$ and the total mass becomes smaller for the same value of the spin (``the balerina effect"), making it challenging to see the effect of the cloud at too small values of $\alpha$. A dedicated study is required to find the optimal value of $\alpha$ and to see whether the effect is observable.

Another subtlety with using gravitational wave signal from compact inspirals to detect the presence of the cloud is that the non-spherical metric perturbation induced
by an infalling compact object may be strong enough to cause a mixing between superradiant and non-superradiant levels and induce the disappearance of the cloud, similar to the effect of axion self-interactions.

To summarize, directly probing the structure of the cloud with extreme mass ratio inspirals is an interesting possibility awaiting for a dedicated theoretical study to decide on whether it is feasible. Another possibility worth exploring are the effects of the cloud on the accretion disk of stellar mass black holes. In this case, the cloud could excite resonant modes of the accretion disk, the so called quasi-periodic oscillations.

\subsection{The QCD axion and superradiance}
\label{QCDaxion}
The QCD axion is the best motivated of all axion-like particles and by itself serves as one of the major motivations for the whole axiverse framework, so let us 
summarize here what range of its parameter space will be probed by on-going and future black hole observations. 

Unlike for other axions, the QCD axion mass $\mu_a$ and decay constant $f_a$ are related to each other by (\ref{QCDmass}). Furthermore, non-perturbative string corrections to the QCD axion potential take the form
\[
V_{string}\simeq \Lambda^4e^{-S}\cos{\phi/f_a}\;,
\]
where the energy scale $\Lambda$ is either Planck or string scale (in exceptional cases it might be suppressed by the SUSY breaking scale, $\Lambda^4\sim M_{Pl}^2F_{SUSY}$  \cite{Svrcek:2006yi}) and the instanton action $S$ in explicit constructions is bounded from above as
\be
\label{weakgravity}
S\lesssim {M_{Pl}\over f_a}\;. \ee
$S$ is close to saturating the above bound if a compactification manifold is neither too anisotropic nor strongly warped. It was suggested \cite{ArkaniHamed:2006dz} that the upper bound (\ref{weakgravity}) follows from very general properties of quantum gravity. However, to the best of our knowledge, there is no specific proposal for the exact  numerical coefficient that
should appear in a conjectured sharp version of (\ref{weakgravity}).

To solve the strong CP problem the instanton action should be sufficiently large, $S\gtrsim 200$. Combined with the above arguments this suggests
that the scale $f_a$ for the QCD axion is unlikely to be significantly higher than $few\times 10^{16}$~GeV or equivalently, that the QCD axion mass is unlikely to be significantly lighter than $10^{-10}$~eV. Also, from a bottom-up perspective, the values of $f_a$ close to the grand unification scale,
corresponding to masses $m_a\sim 3\cdot 10^{-10}$~eV, appear to be well-motivated.

\subsubsection{Black hole spindown and Advanced LIGO}
These arguments motivate thinking of the consequences of superradiance in an axion mass range that is as heavy as possible while still affecting black hole dynamics. Of course, from a purely phenomenological approach any limit on the QCD axion parameters in the high $f_a$ regime are still extremely interesting. The current measurements of black hole masses and spins, presented in Fig.~\ref{fig_summary}, already suggest an {\it upper} bound on the axion decay constant at the level
\be
\label{currentfa}
f_a\lesssim 2\cdot 10^{17}\mbox{ GeV}\;.
\ee
For higher values of the decay constant, {\it i. e.} lighter axion mass, the gap in the upper panel of Fig.~\ref{fig_summary} would shift towards heavier black hole masses and would contain rapidly spinning black holes inside. Of course, at the moment one should consider this bound as indicative. First, the data points in Fig.~\ref{fig_summary} may be subject to significant systematic uncertainties. For instance, the highest spin black hole in Fig.~\ref{fig_summary} is GRS  1915+105, and the model for the soft X-ray spectrum of this object suggests a much smaller value of the spin $a/r_g\approx0.56$~\cite{Blum:2009ez}, than the one presented here (from \cite{McClintock:2009dn}). Second, rapidly spinning black holes in  Fig.~\ref{fig_summary}
may turn out to be young enough to stay in the gap region---for instance, the age of the second fastest spinning black hole in Fig.~\ref{fig_summary} (LMC X-1) is quite short---of order 5 million years \cite{GouMcClintock}.

All these uncertainties will get rectified with more data coming. Given that a black hole produced as a result of the stellar collapse can be as light as $\sim 2M_\odot$,  spin measurements alone can potentially improve the bound (\ref{currentfa}) by a factor of few. Still, the above theoretical arguments suggest that it may be not enough
to discover the QCD axion. A plausible situation could be that the QCD axion is light enough to affect the dynamics of the lightest stellar mass black hole through superradiance, but is still too heavy to produce a noticeable gap in the spectrum of rapidly rotating black holes. This makes it especially important to study other consequences of superradiance that may allow to discover the QCD axion in such a situation.

One exciting possibility to discover the QCD axion is through observing of the gravitational wave signal from superradiating black holes  at Advanced LIGO, as discussed in section~\ref{wavesignature}. Estimates presented there indicate 
that Advanced LIGO may see the transition signal for the QCD axion, although the annihilation signal will probably have too large frequency to be observed at that experiment. As we mentioned, details of the transition signal---such as the strength of the signal at the maximum intensity---may even provide an estimate for the decay 
constant, which would be a further confirmation that the signal is related to the QCD axion, rather than to some other axion-like particle.
In Fig.~\ref{QCDplot} we zoomed in the high mass region of the plot in Fig.~\ref{transignal} relevant for the QCD axion. 
This plot shows only the estimated signal for the $6g\to 5g$ transitions. For heavy axions the higher $l$ levels are likely to be relevant; our estimates show that the corresponding signal is very similar to the one in Fig.~\ref{QCDplot}. Note that we pick $10^4$ second as a coherent integration time for this plot, so that 
the actual reach can be even better for longer integration times.
We see that Advanced LIGO has the potential to probe the most interesting mass range for the QCD axion, and this is even more true for the Einstein telescope.

Coming to other probes,
unfortunately, it is impossible to use  gravitational waves to probe the shape of the QCD axion cloud around stellar mass black holes as it could be done for lighter axions affecting supermassive black holes. However, it is interesting to study whether the QCD axion cloud may resonantly excite certain characteristic  perturbations in the accretion disc that would signal about the presence of the cloud.

\subsubsection{Photon signals: radio waves, $\gamma$- and X-rays}
Finally, there could be QCD axion specific signatures related to the direct coupling of the QCD axion to Standard Model fields. First, there is a coupling (\ref{axphot}) to photons. As we already discussed this coupling converts axions from the superradiant cloud into photon with the rate given by (\ref{agamma}). This conversion rate is too slow to affect the dynamics of superradiance, but it may still provide an observable signal on Earth. The photons in question are almost monochromatic radio waves, and the production rate (\ref{agamma}) translates in the following flux at Earth,
\[
F_{radio}\sim 10^{-14}{\mbox{W}\over\mbox{m}^2}
\l {10^{16}\mbox{ GeV}\over f_a}\r^2 \l{\epsilon \over 10^{-4}}\r
\l
{B\over 10^8\mbox{ Gauss}}\r^2
\l
{1\mbox{ kpc}\over r}
\r^2\;.
\]
 There are several challenges for this signal to be observable, and they all arise because the frequency of the signal is equal to the axion mass, and corresponds to radio waves with wavelengths of order at least few hundred meters. 
The first difficulty arrises because these wavelengths cannot be observed  on the Earth's surface, since the ionosphere is not transparent for radio waves at these low frequencies. However, this problem may be solved by using space- (or Moon-) based radio telescopes. The major challenge, however, is that the electron density  around a black hole
should  be quite small, $n_e\lesssim 100$~cm$^{-3}$, for the signal to be able to escape from the source. It is very hard to find a source satisfying this property, given that one needs substantial magnetic field in the vicinity of the cloud for efficient axion-to-photon conversion. This possibility will be studied in \cite{JMR}.

Probably a more promising possibility is related to the coupling (\ref{axionaction}) of the QCD axion to the QCD instanton number. To observe  the consequences of this coupling would be especially interesting given that  the interaction (\ref{axionaction}) is a genuine footprint of the QCD axion. An intriguing possibility to achieve this is related to the following observation. As follows from the estimate (\ref{varest})  the axion field in the cloud may reach values of order
\be
\label{cloudfield}
\phi\sim2{\alpha\over l} f_a\;,
\ee
{\it i.e.}, the ratio $\phi/f_a$ becomes of order one. For the QCD axion this ratio is nothing else as but the local value of the CP violating $\theta$-parameter in the QCD. QCD properties are rather different at large values of the $\theta$-parameter---for instance, the pion mass is smaller by a factor of order 3 at $\theta=\pi$ as compared to
the $\theta\approx 0$ vacuum, where we live (see, e.g., \cite{Ubaldi:2008nf} for a recent discussion). Given that (multi)pion exchange is one of the dominant forces responsible for
the nuclear binding it is natural to expect that nuclear binding energies change by order one in the regions with $\theta\sim 1$\footnote{Note, that this effect was not taken into account in the analysis of  \cite{Ubaldi:2008nf}.}. 

Consequently, it is natural to speculate that some of the stable nuclei may become unstable  towards disintegration, and gamma- or beta-decay when they enter in the region of the cloud 
as it approaches the maximum size. Even if  (\ref{varest}) somewhat overestimates the maximum $\theta$ in the cloud, it appears very probable that $\theta$ becomes of order one at least during the Bosenova events. The characteristic timescale for the latter is set by the black hole size and is of order $10^{-5}$ seconds. Consequently, it's only strong and electromagnetic nuclear instabilities that have enough time to be important during the Bosenova event.  

Under the optimistic assumption that  an order one fraction of nuclei in the vicinity of the black hole horizon decays and produces $\gamma$-quanta with MeV energies let us estimate the resulting flux of photons at the Earth from the Bosenova event. If the black hole accretes with an Eddington limited rate, the total amount of matter within a distance of order $r_g$ from the black hole horizon can be estimated as $M_{BH}r_g/\tau_E\sim 10^{35}$~GeV$\sim 10^{-22}M_{\odot}$, where $\tau_E$ is the Eddington time (\ref{Eddington}). This may give rise to the emission of order $10^{35}$ photons with MeV energies on a time-scale of order $10^{-5}$ seconds. This is not very much---for a black hole at 10 pc away one would obtain one photon per 10~m$^2$ at the Earth. However, there are several ways the signal can be significantly stronger. First, the major limiting factor in the above estimate is the amount of matter in the vicinity of the black hole horizon. This amount may be roughly 22 orders of magnitude larger immediately after the supernova explosion. Of course, this is a violent event providing lots of radiation by itself, and also immediately after the explosion the metric perturbation due to the surrounding matter is likely to be strong enough to damp the superradiance. However after the environment cleans up a bit this may give rise to a signal many orders of magnitude stronger than in the Eddington regime. 

Also the signal may last significantly longer if, as the estimate (\ref{cloudfield}) suggests, nuclei may get destabilized not only during the
Bosenova collapse, but also when the cloud is still stable. Another possibility for the signal to last longer is for the Bosenova to produce long-lived axion clumps (such as ``pulsons" of \cite{Makhankov:1978rg}) that would be able to escape from the near-horizon region.

 Finally, rather than directly detecting photons one may look for spectral X-ray lines of exotic elements in the vicinity of the black hole, that could have been formed as a result of the nuclear disintegration triggered by the large axion field\footnote{We thank Steve Kahn for pointing this out.}.  
 
\subsection{Accelerated growth of black holes}
\label{bhgrowth}

\begin{figure}[t] 
 \begin{center}
 \includegraphics[width=7in,trim= 0 70 0 70]{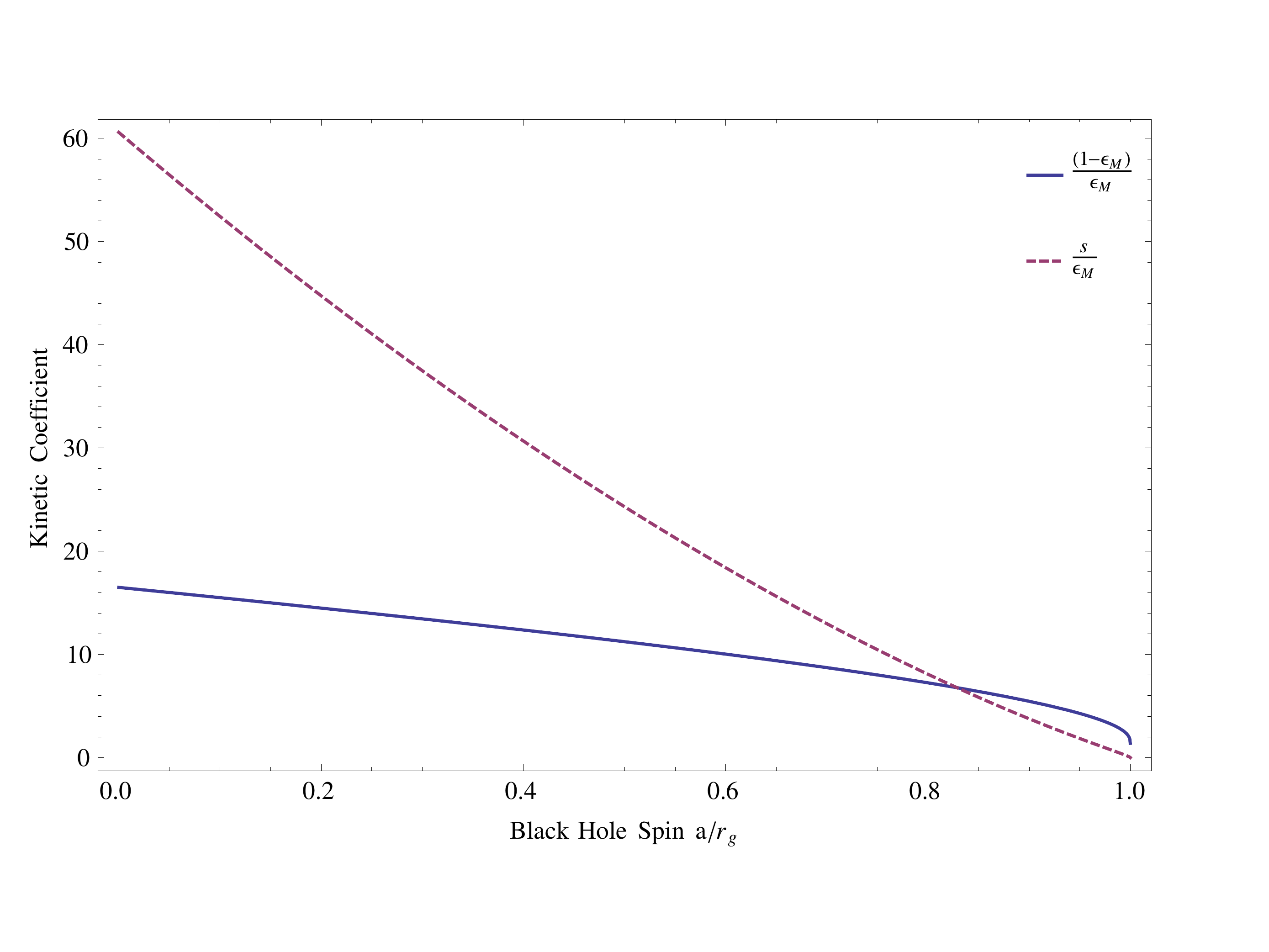}
 \caption{Kinetic coefficients determining the black hole mass (solid line) and spin (dashed line) growth during Eddington limited accretion.}
 \label{shapiro}
 \end{center}
\end{figure}

Yet another indirect consequence of superradiance is that axions, if present, would accelerate the black hole growth. As a toy illustrative example we pick the Eddington saturated thin disc model (see \cite{Shapiro:1983du} for a review and references). In this model the black hole mass evolves according to 
\be
\label{massevol}
{dM\over dt}={1-\epsilon_M(\bar{a})\over\epsilon_M(\bar{a})}{M\over\tau_E}+\dot M_{sr}\;,
\ee
where, the accreting gas is assumed to have zero metallicity,  $\tau_E$ is given by (\ref{Eddington}), and $\epsilon_M(\bar{a})$ is the radiation efficiency for the accrettion---the fraction of the incoming energy that gets radiated away in the process of accretion.
 It depends only on the dimensionless spin-to-mass ratio 
 \[
 \bar{a}\equiv {a\over r_g}\;,
 \]
and in Fig.~\ref{shapiro} we have shown the coefficient $(1-\epsilon_M)/\epsilon_M$ appearing in (\ref{massevol}) as a function of $\bar{a}$. Finally, the $\dot M_{sr}$-term accounts for superradiance, and is given by
\be
\label{Msr}
\dot{M}_{sr}=-\mu_a\sum_i\Gamma_i N_i+\dots\;,
\ee
where the omitted terms are those related to non-linear effects.

The time evolution of the black hole spin in the thin disc model is determined by the following equation
\be
\label{spinevol}
{d\bar{a}\over d t}={s(\bar{a})\over \epsilon_M(\bar{a})\tau_E}+\dot{\bar{a}}_{sr}\;,
\ee
where the ratio $s/\epsilon_M$ as a function of $\bar{a}$ is also shown in Fig.~\ref{shapiro}. As before, the $\dot{\bar{a}}_{sr}$-term describes the effects of superradiance and is equal to 
\be
\label{asr}
\dot{\bar{a}}_{sr}=
-{\mu_a\over M}\sum_i(\alpha^{-1}m_i-2\bar{a})\Gamma_i N_i\;.
\ee
Fig.~\ref{shapiro} now makes the effect of superradiance evident: the solid line there indicates that the mass of a slowly rotating black hole accreting in the Eddington regime grows almost an order of magnitude faster than the mass of a rapidly rotating black hole. 

 \begin{figure}[t] 
 \begin{center}
 \includegraphics[width=6.0in,trim=30 0 0 50]{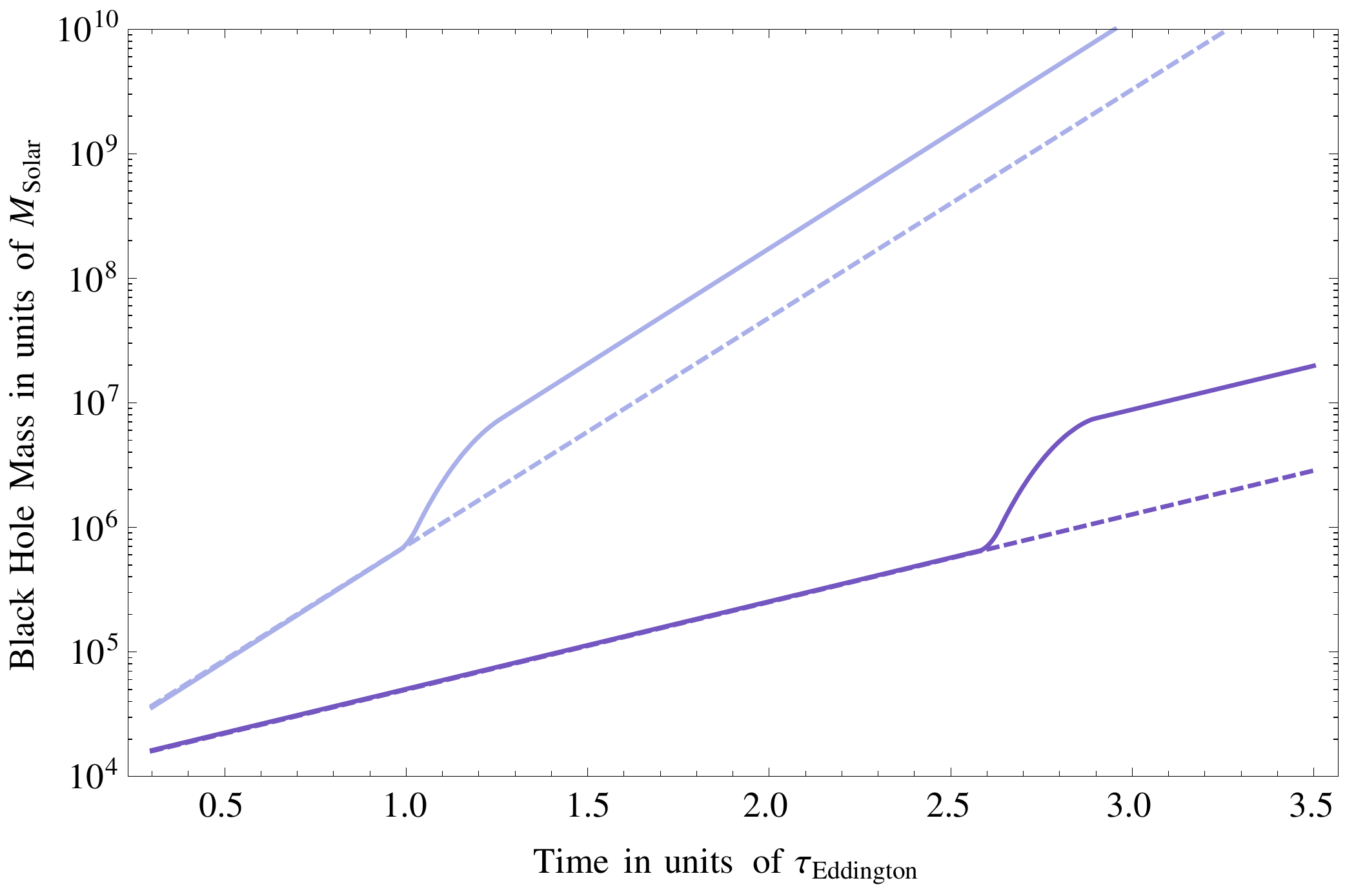}
 \caption{The effect of an axion on the black hole growth history for the simplest  thin disc Eddington limited accretion (lower curves) and for a more realistic model thin disc model of Eddington accretion taking into account result of magnetohydrodynamical simulations (details of both models can be found in \cite{Shapiro:2004ud}).}
 \label{growth}
 \end{center}
\end{figure}
The origin of this effect is easy to understand. For a rapidly rotating Kerr black hole the size of the  last stable orbit is significantly smaller than for a Schwarzschild black hole of the same mass.  As a result the accreting plasma radiates a larger fraction of its rest mass before falling into a rotating black hole. In the Eddington saturated regime the radiation pressure is the main limiting factor for the accretion rate---hence, the accretion proceeds faster for a Schwarzschild black hole, as the solid line in  Fig.~\ref{shapiro} shows.

The dashed line in  Fig.~\ref{shapiro} indicates that a black hole accreting in the Eddington regime rapidly spins up and keeps growing with a high value of spin. Instead, if a black hole is affected by superradiance it follows the Regge trajectory, where its spin can be significantly lower and, as a consequence, the growth rate is much faster. Given that even a single axion affects a large range of black hole masses, superradiance may significantly affect the growth history of supermassive black holes, as illustrated in Fig.~\ref{growth}.

 It is worth noting that quasars hosting $\sim 10^9\;M_{\odot}$ black holes were observed at as high redshifts as $z=6.43$, and some authors think that these observations may present a challenge for the conventional story of black hole growth \cite{Shapiro:2004ud}. Of course, the actual dynamics describing black hole evolution is likely to be significantly more complicted with merger events playing a significant role (see, e.g., \cite{Volonteri:2010wz} for a recent overview), and it is premature to decide whether any new physics, such as axions, is needed to explain the existing observations. However, it is likely that with future X-ray and gravitational wave data as well as with a progress in numerical simulations the evolution history of supermassive black holes will be understood much better, and new physics might eventually be required, especially if quasars with significantly higher redshifts are to be discovered. At any rate, even if a conventional astrophysics is able to explain the data, it is useful to keep in mind that axions, if present, are able to significantly affect the story.

\section{Bookkeeping of anthropic axions}
\label{anthropic}
Before concluding, let us elaborate on one particularly interesting feature of axions in the mass range relevant for the present paper, $\mu_a\gtrsim 10^{-21}$~eV and with a high  symmetry breaking scale, $f_a\sim M_{GUT}$. Namely, the axion abundance relative to baryons 
is given by\footnote{For the QCD axion this formula gets modified  due to the temperature dependence of the axion mass. This is not important for the discussion below.},
\be
\label{abund}
{\Omega_a\over \Omega_b}\simeq 5 \gamma P(\theta_a)\l {\mu_a\over 2.4\cdot 10^{-19}\mbox{ eV}}\r^{1/2}\l{ f_a\over2\cdot 10^{16}\mbox{ GeV}}\r^4\;,
\ee
where $\gamma$ is an order one coefficient different from unity if an axion is heavy enough, so that the effective number of degrees of freedom at the onset of its oscillations is different from the current value;  $\theta_a\equiv \phi(t=0)/f_a$ is an initial axion misalignment angle, and 
\be
\label{apth}
P(\theta_a)\approx \theta_a^2
\ee
for small $\theta_a$ (the shape of $P(\theta_a)$ for general $\theta_a$ can be found, e.g., in Fig.~4 of   \cite{Arvanitaki:2009fg}). 
We see that axions with masses significantly heavier than $\sim 10^{-19}$~eV would produce a contribution to the dark matter density larger than the observed value $\Omega_{cdm}\approx 5\Omega_b$, unless we happened to have an atypically small initial misalignment angle $\theta_a$.

 As was realized long ago \cite{Linde:1987bx}, this does not mean that such axions are necessarily in conflict with the observed value of $\zeta\equiv\Omega_{cdm}/\Omega_b$. Indeed, if inflation lasted sufficiently long (and especially if there were a period of eternal infaltion in the past) an initial misalignment angle $\theta_a$ is a dynamical parameter that varies
in space on scales much longer than the current size of the Universe, so there will always exist regions with sufficiently small value of $\theta_a$ to be in agreement with the observed value of $\zeta$. Still, one may wonder what is the probability for  an observer  in such a Universe to find himself in a region with as small values of
$\zeta$ as we observe.  

For a single  QCD axion this question was addressed a number of times in the past \cite{Tegmark:2005dy,Freivogel:2008qc}. These treatments differ in some details, however they agree that the observed value of $\zeta$ does not appear to be anomalously small. Let us see how this conclusion changes if there are more than one axions with masses  $\gtrsim 10^{-19}$~eV.
Note that depending on which of the parameters are allowed to vary the answer may be more or less sensitive to the uknown details of the statistics of string vacua and to the infamous ambiguities with the probability measure in an eternally inflating Universe.

 We find the approach of \cite{Freivogel:2008qc}---to keep all the parameters apart from $\theta_a$ fixed---the safest from this point of view. In other words, we are restricting to comparing observers with all microphysical parameters the same as ours, but the inflationary  dynamics  automatically produce different initial values of $\theta_a$ for them. In the string landscape the axion abundance is the last parameter that may vary, so this approach is maximally close to the logic applied for predicting the results of a conventional lab experiment---we fix
all particles physics parameters to the known values and see what the dynamics of the system gives us. 
The important difference with a lab experiment is that now we cannot  ignore the selection effects---the formation of observers is impossible in the regions where the dark matter-to-baryon ratio $\zeta$ is either too big or too small. These so called anthropic boundaries were estimated in \cite{Tegmark:2005dy}. Namely, for 
$\zeta\lesssim2.5$ perturbations at the scales close to our galaxy's cease to grow, while at $\zeta\gtrsim 100$ the density of baryons becomes so small that the 
disc fragmentation instability leading to star formation does not develop. The observed value $\zeta\approx 5$ appears to be somewhat too close to the lower end 
of this interval.

 Unfortunately, to quantify whether there is a real problem, we are still left with an ambiguity  related to the choice of the inflationary measure---the problem of comparing numbers of observers
measuring different values of $\zeta$ given that these numbers are infinite in an infinite Universe for any $\zeta$ in the anthropically allowed region.
Following \cite{Freivogel:2008qc}, let us consider what happens with one particular  choice---the causal diamond measure of  \cite{Bousso:2006ev}.
This choice amounts to counting the number of observers in a single Hubble patch of the late time de Sitter evolution. 
Another simplifying assumption of \cite{Freivogel:2008qc} is that the number of observers per baryon is approximately constant for $2.5<\zeta<100$ and
zero otherwise.

The nice feature of the axion setup is that the prior probability distribution for $\zeta$ is known. Indeed, generalizing (\ref{abund}) to the case when more than one 
axion is present  we find  
\be
\label{multabund}
\zeta(\theta_a)=\sum_a c(\mu_a)P(\theta_a)\;.
\ee
The initial values $\theta_a$ is getting set  during inflation when the axion backreaction on the cosmological expansion is negligible, so that the prior distributions
for all $\theta_a$ are flat.
 Then the probability to observe the dark matter-to-baryon ratio smaller than the observed $\zeta=5$ value
 is equal to
\be
\label{probab}
{\cal P}={\cal N}^{-1}{\int_{2.5<\zeta(\theta_a)<5}{\prod_a d\theta_a\over 1+\zeta(\theta_a)}}\;,
\ee 
where a factor $(1+\zeta(\theta_a))^{-1}$ is specific to the causal diamond measure and appears because the total number of baryons within a horizon at the transition to the de Sitter regime is proportional to this factor. The normalization factor ${\cal N}$ is equal to
\be
\label{Norm}
{\int_{2.5<\zeta(\theta_a)<100}{\prod_a d\theta_a\over 1+\zeta(\theta_a)}}\;.
\ee
Expressions (\ref{probab}), (\ref{Norm}) are significantly simplified in the limit when all anthropic axions are sufficiently heavy, so that the approximation
(\ref{apth}) is accurate. In this regime one can get rid of the mass dependence in (\ref{multabund}) by rescaling $\theta_a\to c(\mu_a)^{-1/2}\theta_a$.
As a result, after integration over angular variables in the $\theta_a$ space,  one obtains,
\be
\label{easyprob}
{\cal P}(n)={
\int_{2.5}^5{d\zeta\zeta^{(n-2)/2}\over 1+\zeta}
\over\int_{2.5}^{100}{d\zeta\zeta^{(n-2)/2}\over 1+\zeta}}=0.3,\;0.16,\;0.06,\;0.02,\;0.006,\dots
\ee
where $n$ is the number of axions, and we presented the numerical value of the probability for the first few values of $n$. We see that 
the probability drops exponentially as the number of axions in the anthropic window grows, however, remains high enough for the first few values of $n$.
Clearly, this general trend---that at large number of axions the probability distribution is peaked at the higher anthropic boundary for $\zeta$
 is generic and independent of the choice of the measure. It is just a consequence of a geometrical factor $\zeta^{(n-2)/2}$ in the numerators of integrals in (\ref{easyprob}).
  For instance,
if we droped the $(1+\zeta(\theta_a))^{-1}$ factor and just used the prior probability distribution for $\zeta$ (restricted to the anthropically allowed region), we would get $0.08,\;0.03,\;0.007,\;0.001$ for the first few probabilities. It is worth pointing out that these probabilities are sensitive to the position of the anthropic boundary at large $\zeta$, which is not the case in the presence of a single axion, as pointed out in \cite{Freivogel:2008qc}.

If a sufficiently large number of axions is to be discovered in the anthropic region or, even if, for a single QCD axion, it turns out that a significant fraction of cold dark matter is composed of WIMP's, these probabilities may start being problematically low. However, one should keep in mind that there are lots of uncertainties in the above estimates. Apart from a choice of the inflationary measure, the assumption that the number of observers per baryon is constant over the whole anthropic interval appears to be a vast oversimplification, due to both astrophysical and astrobiological reasons. On the astrophysical side
it is far from clear that the number of stars is proportional to the number of baryons in the whole range $2.5<\zeta<100$. Furthermore, the number of observers  
may  not scale linearly with the number of stars both due to astrophysical reasons, for example due to close encounters, and due to astrobiological, if the early stages of the evolution of life
can be significantly accelerated by the possibility of the transfer of organic molecules (or primitive forms of life) from one stellar system to another (given that the closest known planetary system is just 10 light years away this possibility is neither necessarily hypothetical nor untestable).

To summarize, we see that at the current stage of affairs there is no reason to be discouraged on the possibility of a discovery of  multiple anthropic axions with astrophysical black holes observations. Conversely, if several anthropic axions were to be discovered (or even a single one if WIMPs constitute a significant fraction of dark matter) this will provide us with serious motivation to scrutinize how the number density of observers depends on the  baryon-to-dark matter ratio.

\section{Conclusions}
\label{fuuh}

We hope to have convinced the reader that black hole superradiance for axions is an extremely rich phenomenon that has good chances to be observed in near future measurements of black hole properties. Ongoing black hole spin measurements may trace gaps or Regge trajectories in the spectrum of rapidly spinning black holes. Advanced LIGO may observe gravitational wave signals from the QCD axion cloud around stellar mass black holes as far as the Virgo cluster for masses down to $10^{-10}$ eV which correspond to an axion decay constant close to the grand unification scale. In a more distant future, gravitational waves may be observed for supermassive black holes at lower frequencies by experiments such as LISA and AGIS. The low frequency gravitational wave detectors may also see the effect of the cloud on the waveforms during extreme mass ratio inspirals. Finally, for the QCD axion the superradiant cloud might also give rise to direct photon signals. 



In this paper our main goal has been to develop a general intuition about superradiance development and its consequences without going into an extensive numerical work. Given the richness of the system it seems inevitable that detailed numerical simulations will be required in the future to obtain accurate quantitative predictions. This is especially important for predicting the strength, duration and precise waveform of gravitational wave signals from superradiant clouds. Simulations are needed both for the accurate prediction of superradiance rates, which to large extent has already been accomplished in \cite{Dolan:2007mj}, and most importantly to get an accurate description of the superradiant cloud including axion self-interactions. 
It's worth stressing, however, that some of our result are very robust. For instance, the black hole Regge trajectories of Fig.~\ref{fig_summary}
are mostly determined by the basic superradiance condition (\ref{Omegacond}) and do not depend on the above uncertainties (apart
from the left most
declining segments of these curves; however, in that region even larger uncertainties are likely to come from variations in the accretion rate for different black holes).  

Of course,  we expect also other qualitative results obtained here to reproduce well the gross features of the system, although given its richness more surprises are possible. Most importantly, our results appear encouraging for prospects to observe superradiance with future astrophysical data, and this justifies further theoretical efforts for better understanding of this fascinating process.

\section{Acknowledgements}
We thank Savas Dimopoulos, John March-Russell and Nemanja Kaloper for collaboration at all stages of the project.
We also thank Tom Abel, Andrei Frolov, Gregory Gabadadze, Andrei Gruzinov, Peter Graham, Steve Kahn, Mark Kasevich, Andrew MacFadyen, Sergei Sibiryakov, Steve Shenker, David Shoemaker and Bob Wagoner for useful discussions.

\section*{Appendix A: Calculating the tunneling exponent (\ref{Integral})}
 
 In principle, it is straightforward to calculate the tunneling integral $I$ numerically, however, let us also describe an anlytical method that works for near-extremal black holes,
 $a\approx r_g$ and for the frequency $\omega$ right at the boundary of the superradiance region $\omega=m w_+$. 
 
 The latter condition has the following meaning.
As we see from  Fig.~\ref{rates}, each of the superradiant levels has a maximum rate close to the boundary of the superradiant region. Of course, exactly at the boundary the width of the level becomes zero. However, it is the pre-factor in (\ref{GammaWKB}) that turns zero at   $\omega=m w_+$, while the exponent $I$ just passes smoothly through that point. Consequently, by calculating $I$ at $\omega=m w_+$ we will find the upper envelope of the family of superradiant rates for different levels. Related to this, we will also set $l=m$, because this corresponds to the fastest superradiance rate at any $\alpha$. Finally,  we also set $\mu_a=\omega$, which given the above assumptions corresponds to taking $l=2\alpha$. This should be a good approximation given that superradiant levels are close to be non-relativistic. Note that for the fastest level the radial number $n$ also grows with  $l$, so that the upper bound (\ref{alpha_bound}) never gets saturated. All these assumptions were also made in \cite{Zouros:1979iw}.
 
 The simplification at $\omega=mw_+$ is that the location of $r_1$ is known, namely $r_1=r_+$, for this choice of parameters. This fact is straightforward to check explicitly using (\ref{potential}). It is also easy to understand intuitively---the only way for the tunneling rate (and, consequently, for the imaginary part of an eigenfrequency $\omega$) to vanish is for the tunneling to be ``kinematically forbidden", and this is exactly what happens if $V(r_+)=0$. For all other values
 of $\omega$ the potential at the horizon is negative, $V(r_+)=-(\omega-m w_+)^2$ and the imaginary part is non-zero.   
 
 \begin{figure}[t] 
 \begin{center}
 \includegraphics[width=7in,trim=0 50 0 100]{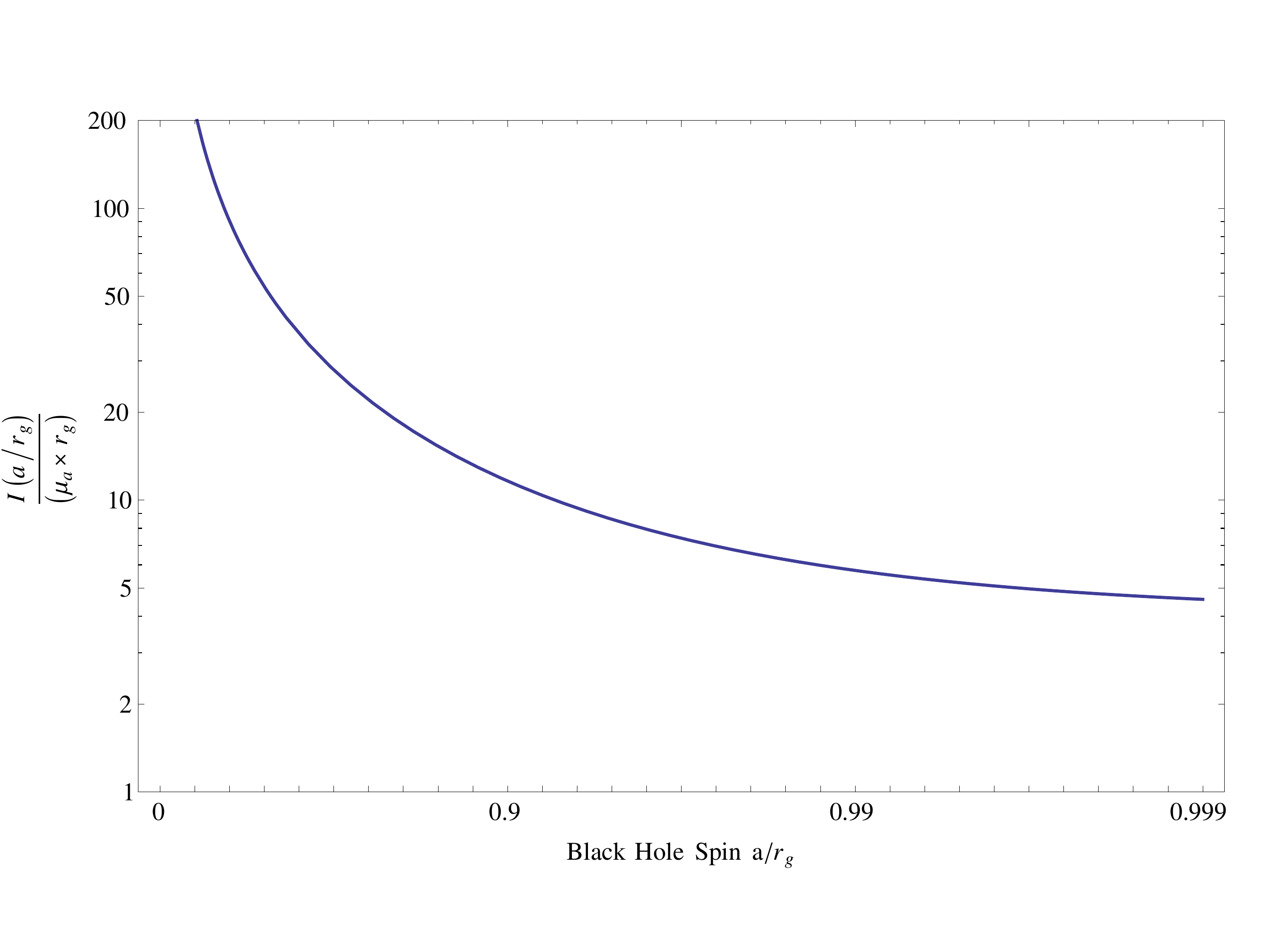}
 \caption{The tunneling exponent as a function of the black hole spin.}
 \label{IWKB}
 \end{center}
\end{figure}

 A further simplification happens at $a=r_g$---in this case the second turning point coincides with the horizon, $r_2=r_+$ for $l>2$. This implies that at $\omega=mw_+$
  one cannot  set $a=r_g$ before performing the integral in (\ref{Integral}). However, this makes the integration simple.
Indeed at $\omega=mw_+$ we can write
  \be
  \label{vdef}
  V(r)=(r-r_+)(r_2-r)v(r)\;,
  \ee
  where in the limit $a\to r_g$ the function $v(r)$ has a finite non-vanishing limit in the whole interval $(r_+, r_2)$. Hence at $\omega=mw_+$,  we can write the integral (\ref{Integral}) in the limit $a\to r_g$ as
  \be
  \label{Int1}
  I=v(r_+)^{1/2}(r_+^2+a^2)\left.\int_{r_+}^{r_2}dr\sqrt{r_2-r\over r-r_+}{1\over r-r_-}\right|_{a\to r_g}=2\pi v(r_g)^{1/2}r_g^2\left.\l\sqrt{r_2-r_-\over r_+-r_-}-1\r\right|_{a\to r_g}\;.
  \ee
 Now, taking the second derivative of (\ref{vdef}) with respect to $r$ and setting $a=r=r_g$  we obtain,
 \be
 \label{vrg}
 \left.v(r_g)\right|_{a=r_g}=-{1\over 2}\d^2_rV(r_g)|_{a=r_g}={\alpha(\alpha-1)\over 2 r_g^4}\;.
 \ee
 Similarly, taking the mixed second derivative of (\ref{vdef}) with respect to $r$ and $a$ we obtain,
 \be
 \label{dra}
\sqrt{r_g^2-a^2}\left. \d_a r_2\right|_{a=r_g}=\left.\sqrt{r_g^2-a^2}\d_a\l{\d_r V(r_g)\over g(r_g)}+r_+\r\right|_{a=r_g}={3\alpha+1\over 1-\alpha}
 \ee
 Equivalently, (\ref{dra}) implies that at $a\approx 1$ one has
 \be
 \label{r2appr}
 r_2=r_g+{3\alpha+1\over \alpha-1}\sqrt{r_g^2-a^2}+{\cal O}(r_g^2-a^2)\;.
 \ee
 Finally, by plugging (\ref{vrg}) and (\ref{r2appr}) into (\ref{Int1}), we obtain the following answer for the tunneling integral in the extremal Kerr geometry
 \be
 \label{IntegralW}
 I=\pi \l 2\alpha-\sqrt{2\alpha(\alpha-1)}\r\;.
 \ee

 In principle, we can continue as above and work out higher order terms in the $(1-a)$-expansion to arrive at the approximate analytical WKB formulaes for superradiant rates in the near-extremal case. Instead, in Fig.~\ref{IWKB} we present the result of a numerical integration of (\ref{Integral}) as a function of the black hole spin.

\end{document}